\documentclass[pdflatex,sn-mathphys-num]{sn-jnl}

\usepackage{graphicx}%
\usepackage{multirow}%
\usepackage{amsmath,amssymb,amsfonts,physics}%
\usepackage{amsthm}%
\usepackage{mathrsfs}%
\usepackage[title]{appendix}%
\usepackage{xcolor}%
\usepackage{textcomp}%
\usepackage{manyfoot}%
\usepackage{booktabs}%
\usepackage{algorithm}%
\usepackage{algorithmicx}%
\usepackage{algpseudocode}%
\usepackage{listings}%
\usepackage{siunitx}

\theoremstyle{thmstyleone}%
\theoremstyle{thmstyletwo}%

\theoremstyle{thmstylethree}%

\begin{document}

\title[Article Title]{Automated Discovery of Operable Dynamics from Videos}


\author[1]{\fnm{Kuang} \sur{Huang}}\email{kh2862@columbia.edu}
\equalcont{These authors contributed equally to this work.}

\author[2]{\fnm{Dong Heon} \sur{Cho}}\email{dongheon.cho@duke.edu}
\equalcont{These authors contributed equally to this work.
\begin{center}
\textcolor{orange}{\url{http://generalroboticslab.com/SmoothNSV}}
\end{center}}

\author*[2,3,4]{\fnm{Boyuan} \sur{Chen}}\email{boyuan.chen@duke.edu}

\affil[1]{\orgdiv{Department of Applied Physics and Applied Mathematics}, \orgname{Columbia University}, \orgaddress{\street{500 W. 120th Street}, \city{New York}, \postcode{10027}, \state{NY}, \country{USA}}}

\affil[2]{\orgdiv{Department of Computer Science}, \orgname{Duke University}, \orgaddress{\street{308 Research Dr.}, \city{Durham}, \postcode{27705}, \state{NC}, \country{USA}}}

\affil[3]{\orgdiv{Department of Mechanical Engineering and Material Science}, \orgname{Duke University}, \orgaddress{\street{121 Hudson Hall}, \city{Durham}, \postcode{27708}, \state{NC}, \country{USA}}}

\affil[4]{\orgdiv{Department of Electrical and Computer Engineering}, \orgname{Duke University}, \orgaddress{\street{100 Science Dr.}, \city{Durham}, \postcode{27708}, \state{NC}, \country{USA}}}


\abstract{Dynamical systems form the foundation of scientific discovery, traditionally modeled with predefined state variables such as the angle and angular velocity, and differential equations such as the equation of motion for a single pendulum. We introduce a framework that automatically discovers a low-dimensional and operable representation of system dynamics, including a set of compact state variables that preserve the smoothness of the system dynamics and a differentiable vector field, directly from video without requiring prior domain-specific knowledge. The prominence and effectiveness of the proposed approach are demonstrated through both quantitative and qualitative analyses of a range of dynamical systems, including the identification of stable equilibria, the prediction of natural frequencies, and the detection of chaotic and limit cycle behaviors. The results highlight the potential of our data-driven approach to advance automated scientific discovery.}

\keywords{Dynamical Systems, Machine Learning, Representation Learning, AI4Science}



\maketitle





\section{Introduction}\label{sec1}

Dynamical systems drive the discovery of physical laws from natural phenomena across scientific and engineering disciplines \citep {strogatz_nonlinear_2015}. By distilling complex observations into key variables and equations, they provide compact and operable representations that not only enable predictive simulations, but more notably, deliver analytical insights.

For instance, in the case of a swinging pendulum, variables such as angle and angular velocity, along with the equation of motion acting on these variables, allow for the prediction of system dynamics from any initial state, as well as the identification of equilibrium states and characterization of periodic motions. This paradigm forms the foundation of modern science, ranging from classical mechanics to fluid dynamics and quantum mechanics.

However, deriving such representations for new systems remains a labor-intensive and time-consuming process. Historically, the analysis of even a simple system like a swinging pendulum, through the equation of motion expressed as a differential equation of the pendulum angle and angular velocity, has required centuries of work by physicists \cite{Palmieri2009-dx,Bell1941-px}. For more complex and high-dimensional systems, in fields such as biology, the formulation and verification of representative mathematical models still require substantial effort and domain-specific knowledge because first-principles equations have yet to be discovered \cite{Luo2021-gk,Martinez-Calvo2022-xe}.

Although many advances in sensing and experimental techniques have led to an abundance of high-quality data, identifying a compact and operable state space for unknown systems remains a predominantly manual and slow task. With the recent improvements in computational power and machine learning, utilizing AI to facilitate the scientific discovery process starts to show increasing promise \cite{Baig2023-jw,schneider2023learnable,ma2024llmsimulationbileveloptimizers}. 


Modern deep learning techniques are effective at processing complex high-dimensional data and performing predictive simulations ranging from weather forecasts to protein structure predictions \cite{ALQURAISHI20211,brunton2020machine,Jumper2021-wd, Azizzadenesheli2024-ga}. However, while these approaches have achieved remarkable prediction accuracy, progress on the equally important goal of data-driven dynamical systems research, namely, automatically extracting insights from observational data, remains rather limited. In other words, current data-driven methods tend to emphasize \underline{\textit{forecasting over interpretability}}, often producing results that are difficult to integrate into the broader framework of scientific understanding.

The outcomes from most deep learning frameworks take the form of high-dimensional signal predictions \cite{kovachki2021neural,li2021fourierneuraloperatorparametric,Gao2024-ej} or learned latent features that are difficult to perform further calculus-based analysis or quantitative reasoning \cite{lusch2018deep,seidman2022nomadnonlinearmanifolddecoders,seidman2023variationalautoencodingneuraloperators,Du2024-tb}. This stands in stark contrast to classical dynamical systems, where physically grounded variables and equations enable concrete analysis. As a result, most deep learning models cannot deliver further actionable scientific conclusions or principles, limiting their utility in discovery-oriented domains.

To address this limitation, our research shifts the focus from purely predictive modeling to interpretable system representation. Unlike prior approaches, we propose a framework that, from raw video recordings of dynamical systems, automatically produces system representations that are not only compact but also operable -- enabling the discovery of system properties such as equilibrium states, natural frequencies, and chaotic regimes. Our framework allows scientists to analyze system dynamics directly from raw observational data with minimum assumptions to accelerate scientific discovery.

Pioneering efforts in this direction have focused on symbolic equation discovery \cite{bongard2007automated, schmidt_distilling_2009, schmidt_automated_2011, brunton2016discovering, cranmer2020discovering}, which have been successful in recovering governing equations from low-dimensional \cite{brunton2016discovering, cranmer2020discovering, Kaheman_2020} and some high-dimensional observations \cite{champion2019data,bakarji2022discoveringgoverningequationspartial}. However, these approaches often rely on many system-specific assumptions, such as prior knowledge of state dimensionality or a fixed library of nonlinear terms. Hence, they are typically limited to structured datasets such as spatiotemporal fields with known physical variables, where high-order derivatives can be estimated reliably to guide the equation discovery process. These constraints significantly reduce their applicability to unstructured or novel systems where such assumptions cannot be made a priori.

\begin{figure}[H]
    \centering
    \includegraphics[width=1\textwidth]{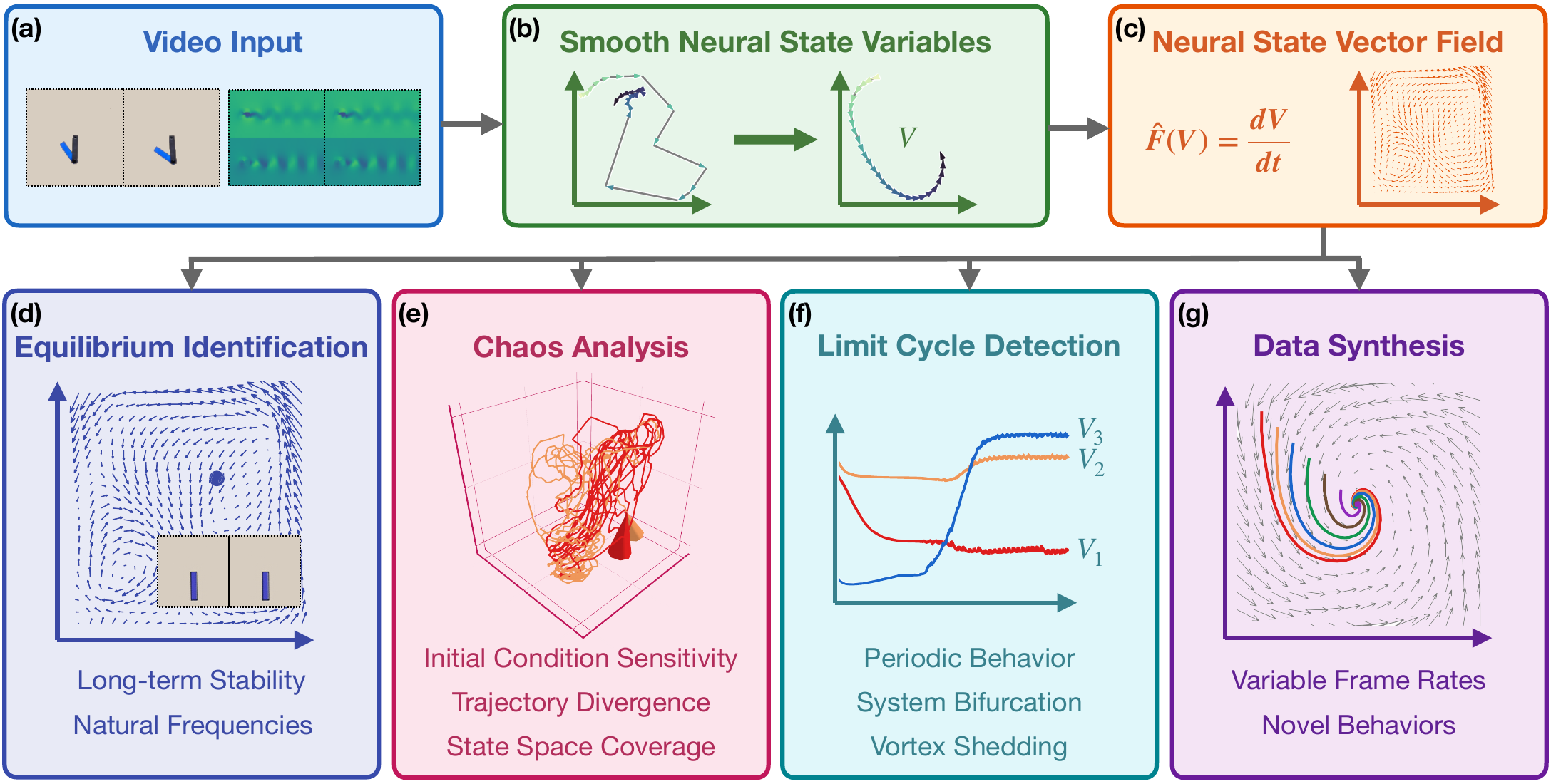}
    \caption{The pipeline of our method to extract smooth neural state variables and neural vector field from videos. (a)~Our framework automatically extracts compact and operable representations directly from observational data, provided as video frames. (b)~Our minimally intrusive smoothness constraints enforce the neural state variables to exhibit smooth trajectories. (c)~We trained an additional neural network $\hat{F}$ to represent the neural state vector field that describes the system dynamics of the discovered smooth neural state variables. Our discovered smooth neural state variables and neural state vector fields allow various scientific analyses, as exemplified in the following sub figures. (d)~The identified stable equilibrium state and the decoded images of the single pendulum are marked within its neural state vector field; (e)~Indicating chaotic behavior, two nearly identical initial states of the double pendulum exhibit diverging smooth neural state variable trajectories; (f)~The developing oscillations in a smooth neural state variable trajectory indicate that the cylinder wake system's dynamics is attracted to a limit cycle, corresponding to the laminar periodic vortex shedding of the flow; (g)~The damped neural state vector field of the spring mass system push integrated trajectories towards the stable equilibrium. More results on all systems can be seen in later sections.
    }\label{fig:pipeline}
\end{figure}

In contrast, our framework is the first to extract compact and operable dynamical representations directly from high-dimensional raw data, such as video frames, without requiring system-specific assumptions. Rather than assuming known measurable physical variables \cite{Narasingam2018-jl, moore2024automatedglobalanalysisexperimental} or state dimensionality \cite{Conti_2023, conti2024venivindyvicivariational}, our method learns a set of state variables and a vector field governing their temporal evolution solely from videos. Together, they form a minimal package that approximates the system dynamics and is operable for calculus-based analysis, enabling the integration of the vector field to reconstruct system trajectories and suggest potential system properties. 
As shown in Figure~\ref{fig:pipeline}, our framework supports automated human-interpretable system analysis directly from high-dimensional data streams.

A key strength of our method lies in its independence from system-specific assumptions. For example, frameworks based on Hamiltonian \cite{greydanus2019hamiltonian} or Lagrangian \cite{cranmer_lagrangian_2020} mechanics fail to model dissipative systems like damped oscillators. Similarly, linear decomposition techniques such as Dynamic Mode Decomposition (DMD) \cite{schmid2010dynamic, Kutz2016-vm} or Proper Orthogonal Decomposition (POD) \cite{berkooz1993proper}, as well as interpretable reduced-order models that impose linear dynamics \cite{kaltenbach2023interpretablereducedordermodelingtimescale}, often fall short in capturing nonlinear behavior. Other methods enforce strong inductive biases, like manifold separation \cite{Floryan2022-xx} or rigid-body assumptions \cite{zhu2025continuitypreservingconvolutionalautoencoderslearning}, which limit flexibility and may distort the true system dynamics. Indeed, recent studies show that imposing physics-inspired priors can degrade predictive performance in complex systems \cite{botev2021priorsmatterbenchmarkingmodels}.

Instead, our framework requires no prior knowledge of the system's dynamical properties, key variables, or their derivatives. By deriving representations directly from unprocessed data, our method provides a robust and generalizable foundation for analyzing complex systems, free from potentially unavailable or incorrect system-specific assumptions. To achieve such compact and operable representations, our approach compresses high-dimensional raw data into a minimal set of state variables while enforcing only the essential structural constraints, such as smoothness of system trajectories and state space occupancy, through enforced regularization techniques and targeted training strategies.

We demonstrate the effectiveness of our method across four benchmark systems spanning classical mechanics and fluid dynamics. Without assuming system-specific prior knowledge, our approach supports rich dynamical analyses that are traditionally enabled by governing equations: identifying equilibrium states, estimating natural frequencies of linear and nonlinear oscillators, distinguishing chaotic from regular behavior, and detecting bifurcation from steady-state flows to periodic vortex shedding. 
Although analytical justification and verification still require involvement from human scientists, these analyses serve as useful starting points for further analysis for scientific discovery. Our results highlight the potential of AI-assisted frameworks to augment scientific reasoning, offering a new paradigm for data-driven discovery that bridges predictive accuracy and interpretability.

\section{Results}\label{sec:results}

\subsection{Datasets}
In this section, we introduce the four systems chosen to demonstrate our method. As shown in Figure~\ref{fig:datasets}(a), these systems include: a spring mass, a single pendulum, a double pendulum, and a cylinder wake system describing fluid flow passing around a cylinder. 
The spring mass and single pendulum systems are canonical examples in classical mechanics, each characterized by two state variables. The spring mass system is a linear oscillator with a constant frequency, and the single pendulum system is a nonlinear oscillator with amplitude-dependent frequencies. 
The double pendulum is a classic example of a complex nonlinear system that exhibits both regular and chaotic behavior. The system requires four state variables to fully describe its state, and unlike the spring mass and single pendulum, its dynamics cannot be expressed with closed-form solutions.
Moreover, our dataset consists of real-world video recordings of a physical double pendulum, for which the exact equations of motion are unavailable, and analyses must rely on approximate physical parameters and idealized conditions such as no frictional forces. Despite these challenges, our method is capable of deriving the governing dynamics entirely from raw visual observations, without any access to physical measurements or prior assumptions on system dynamics. The discovered dynamics and corresponding analyses align closely with those from classical mechanics \cite{Levien1993-pz}.
Finally, the cylinder wake system illustrates the applicability of our method beyond mechanical systems. This well-studied fluid dynamics example exhibits a bifurcation from a laminar steady wake to laminar periodic vortex shedding as the Reynolds number increases \cite{brunton2016discovering,duvsek1994numerical,barkley2006linear,noack2003hierarchy}. Using our method, we discovered that only three state variables, instead of the entire velocity field that includes velocities at all spatial locations, can describe the system state at a given time. This finding is consistent with the results discovered in refs.~\cite{brunton2016discovering,duvsek1994numerical} using mathematical and numerical analysis. For more details on these four systems, please refer to Section~\ref{appendix:dataset}  of the \nameref{appendix}. 

\subsection{Discovering smooth neural state variables and neural state vector fields}

Our method distills video recordings of the studied systems into a state space of $d$ variables, where $d$ is the system's intrinsic dimension - the minimum number of variables to fully describe the system. Provided with sequential data $\{\mathbf{X}_0, \mathbf{X}_{\Delta t}, \mathbf{X}_{2\Delta t}, \cdots\}$ consisting of video frames $\mathbf{X}_t \in \mathbb{R}^{128\times128\times3}$ sampled at a constant interval $\Delta t$, our method compresses the corresponding high dimensional state representations $\{\mathbf{C}_0, \mathbf{C}_{\Delta t}, \mathbf{C}_{2\Delta t}, \cdots\}$ into trajectories $\{\mathbf{V}_0, \mathbf{V}_{\Delta t}, \mathbf{V}_{2\Delta t}, \cdots\}$, where $\mathbf{C}_t = (\mathbf{X}_t, \mathbf{X}_{t+\Delta t}) \in \mathbb{R}^{128\times256\times3}$ denotes the system state at time $t$ in the form of concatenated image data composed of two consecutive frames and $\mathbf{V}_t\in\mathbb{R}^d$ represents the respective compressed state variables, by eliminating redundant visual information while retaining key dynamics. Moreover, our method imposes smoothness constraints to ensure gradual changes from $\mathbf{V}_t$ to $\mathbf{V}_{t+\Delta t}$ for small $\Delta t$, reflecting the continuous nature of the system dynamics.


The resulting state variables $\mathbf{V}_t$, which satisfy both compactness and temporal smoothness, are termed \emph{smooth neural state variables}. Building upon the trajectories encoded in the smooth neural state variable space, we can extract first-order derivatives and define a vector field over the state space. The derived vector field, termed the \emph{neural state vector field}, endows every point in the state space with the system's dynamic information, even in regions where no data are observed. Together, these components form a fully differentiable representation that enables calculus-based analysis, mirroring the structure of classical equations of motion using physical variables such as angle and angular velocity.
Importantly, our approach introduces only minimal and non-intrusive constraints, which make no assumptions about the symbolic form of the dynamics, the type of system (e.g., Hamiltonian, Lagrangian, rigid-body), or even the dimensionality of the state space.
The entire pipeline is automated, as shown in Figure~\ref{fig:pipeline}(a)-(c).


To achieve this transformation, we utilized two stacked auto-encoders which were trained separately, building upon a recent work \cite{chen_automated_2022}. 
This two-step design is motivated by two reasons: (1) directly compressing high-dimensional video into low-dimensional variables often sacrifices prediction accuracy, and (2) our method can discover the dimensionality $d$ automatically, without prior knowledge.
While the resulting \emph{neural state variables} can produce reliable long-term predictions~\cite{chen_automated_2022}, in their base form, they lack smoothness and are unsuitable for calculus-based system analysis.  In this work, we introduce additional constraints to enforce temporal smoothness in $\mathbf{V}_t$, relying only on the reasonable assumption that the system evolves continuously over time. These smoothness constraints limit large jumps between consecutive states, as demonstrated in Figure~\ref{fig:pipeline}(b). 
Additionally, we applied further space-filling constraints to encourage these trajectories to spread throughout the state space and prevent the smoothness constraint from overpowering the optimization procedure, which can lead to the collapse of all trajectories into a narrow region of the state space. For more details on this collapsing behavior, please refer to Section~\ref{appendix:enforcing_smoothness} of the \nameref{appendix}.
These two constraints define an operable representation space that is both compact and structured, while avoiding strong inductive biases that could distort the true dynamics. 
For more details on the compression method and regularization constraints, please see \nameref{sec:method:variables} and \nameref{sec:method:smoothness} of \nameref{sec:methods}.

The derivatives $\frac{\dd\mathbf{V}_t}{\dd t}$ of the resulting smooth neural state variables are well-defined, depending only on the current state $\mathbf{V}_t$. Thus, the second stage of our method aims to extract an accurate representation of the vector field $\hat{F}:\,\mathbb{R}^d\to\mathbb{R}^d$ such that:
\begin{align}\label{eq:vector_field}
    \frac{\dd\mathbf{V}_t}{\dd t} = \hat{F}(\mathbf{V}_t),
\end{align}
as shown in Figure~\ref{fig:pipeline}(c).
We term this vector field as the system's \emph{neural state vector field}, which was implemented as a neural network and trained to predict neural state variable trajectories by integrating Equation~\eqref{eq:vector_field}. For more details on the implementation and training of the neural state vector field, please refer to \nameref{sec:method:vectorfield} of \nameref{sec:methods}. We note that any insights into potential system properties from this learned $\hat{F}$ approximating the ground truth dynamics rely on the structural stability of the system and the adequate spatial temporal coverage of the video data. Rigorously speaking, analyses drawn from such approximated dynamics only serve as an initial guide for further evaluation. Further verification and analytical justification on the global dynamics are required to fully substantiate any derivable conclusions.


\begin{figure}[H]
    \centering
    \includegraphics[width=1\textwidth]{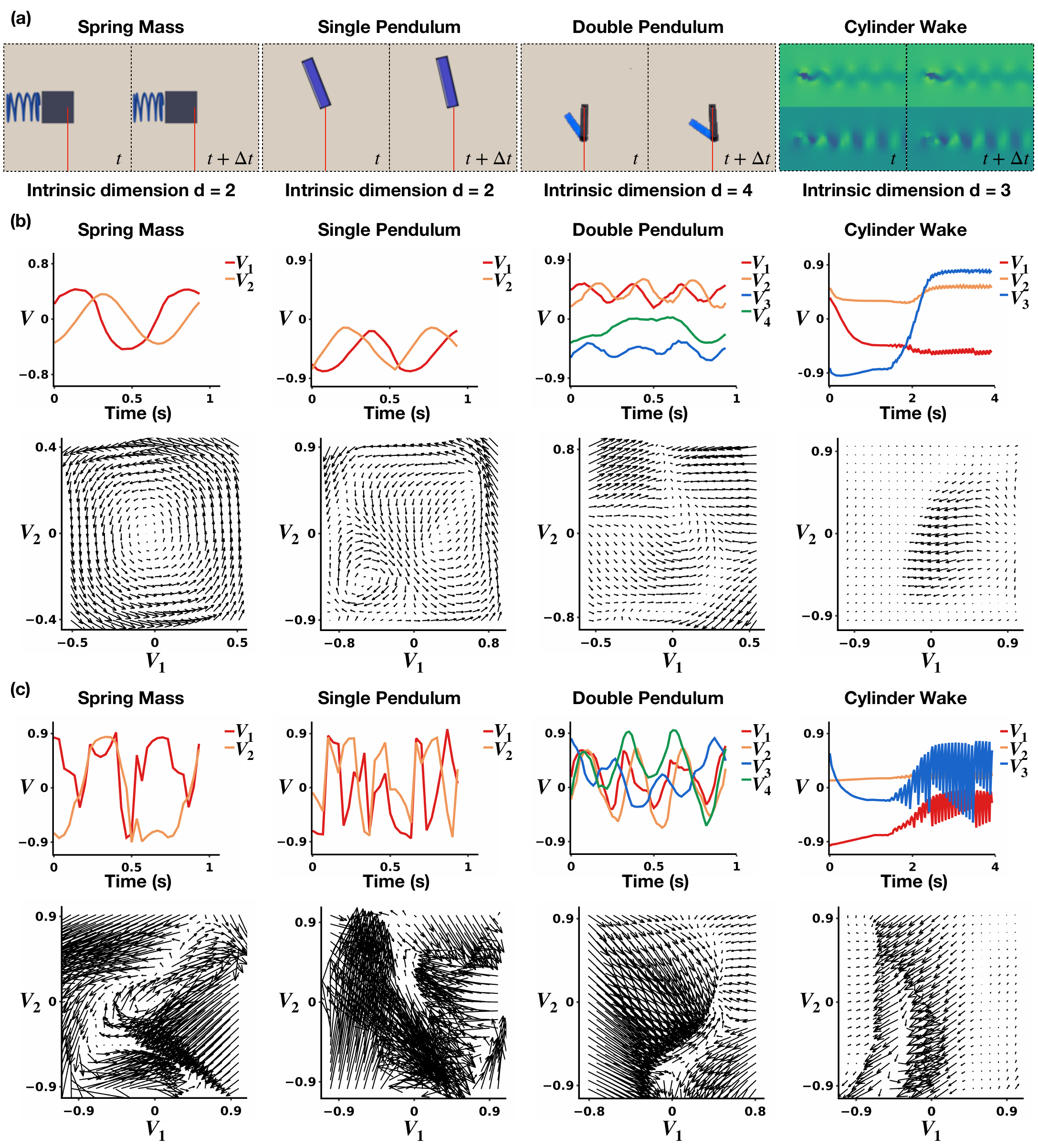}
    \caption{Sample video frames and discovered representations. (a) Sample images of the four studied systems are shown with their respective intrinsic dimension: the spring mass ($d=2$), the single pendulum ($d=2$), the double pendulum ($d=4$), and the cylinder wake ($d=3$). Red vertical lines indicate zero positions or angles. (b) Visualizations of the respective discovered smooth neural state variables and neural state vector fields for the four studied systems demonstrate our framework's ability to extract smooth state trajectories that follow well defined continuous dynamics. (c) The baseline neural state variables and neural state vector fields trained on baseline neural state variables for the four systems demonstrate the highly disorganized state space when no regularization constraints are enforced. For the double pendulum and cylinder wake systems, which have intrinsic dimensions greater than two, we show the neural state vector field in the $V_1$ and $V_2$ dimensions.
    }\label{fig:datasets}
\end{figure}

Figure~\ref{fig:datasets}(b) shows trajectories of our discovered smooth neural state variables and the respective neural state vector fields for the four systems shown in Figure~\ref{fig:datasets}(a). For comparison, we also show in Figure~\ref{fig:datasets}(c) the trajectories and neural state vector fields for baseline neural state variables trained without applying our proposed constraints. We observe that smooth neural state variables demonstrate smoother trajectories and more coherent neural state vector fields. In contrast, the baseline neural state variables' trajectories are frequently disrupted by large jumps and their respective vector fields display abrupt variations. This representation consisting of smooth neural state variables and the neural state vector field forms the foundation for downstream scientific analysis demonstrated in the following sections.

\subsection{Near-equilibrium analysis}\label{sec:equilibrium}

Equilibrium states, which are stationary solutions to a dynamical system, are fundamental for its analysis - they organize the system's phase space and govern its long-term dynamics. 
Moreover, the system's near-equilibrium behavior, such as its stability under small perturbations, is crucial to understanding the system.
In particular, near a stable equilibrium state, the system's dynamics are trapped in the near-equilibrium region where their respective gradients with respect to time remain small, reaching zero at the equilibrium state. This property allows for accurate linear approximations of the dynamics near such states, which can reveal further insight thanks to their simplification of the system.
For example, in the single pendulum system, the state where the pendulum hangs downward with zero velocity is a stable equilibrium state. Releasing the pendulum near this equilibrium state causes it to undergo small oscillation while remaining close to the equilibrium state. The frequency of the oscillation, termed the \emph{natural frequency}, is a fundamental physical property determined solely by the pendulum's length and the gravitational constant. The linearized dynamics in the near-equilibrium region can enable the extraction of the natural frequency.
Therefore, stable equilibrium states offer a natural starting point for analyzing a system's phase space and key properties. 

However, performing near-equilibrium analysis directly with raw observations is challenging. The equilibrium state may not appear in the training data, and high-dimensional or non-smooth latent representations lack a principled notion of distance or continuity to define the near-equilibrium region. Compounding these challenges, applying linearization to derive fundamental physical properties requires computing derivatives of the system's dynamics, which is intractable from discrete data alone. Our smooth neural state variables and neural state vector field address these challenges by providing a smooth, low-dimensional, and fully differentiable representation of the system. This enables derivative-based analysis in the state space through implicit interpolation learned from training data, and supports both stability assessment and natural frequency estimation.

In this section, we demonstrate that our operable representations can be used to discover stable equilibrium states of a system and offer its linearized approximations in the near-equilibrium region. First, we identified equilibrium states by utilizing the neural state vector field $\hat{F}$ and solving for $\vb{V}_{\mathrm{eq}}$ such that:
\begin{align}
    \hat{F}(\vb{V}_{\mathrm{eq}}) = 0.
\end{align}
Given the identified equilibrium states, we further determined their stability through empirical methods. According to the definition of Lyapunov stability \cite{strogatz_nonlinear_2015}, an equilibrium state $\vb{V}_{\mathrm{eq}}$ is stable if for all $\epsilon > 0$, there exists a $\delta > 0$, such that if $\|\vb{V}(0) - \vb{V}_{\mathrm{eq}}\| < \delta$, $\|\vb{V}(t) - \vb{V}_{\mathrm{eq}}\| < \epsilon$ holds for all $t > 0$. 
By integrating the neural state vector field to generate system trajectories, we can numerically assess the stability of the identified equilibrium states by locally evaluating nearby dynamics over a finite horizon. We generated multiple trajectories starting from varying distances to the identified equilibrium state and observed if there exists such a $\delta$ in which all trajectories that begin from initial states within a $\delta$-neighborhood of the equilibrium state remain within a $\epsilon$-neighborhood of the equilibrium state. This process was repeated for a collection of $\epsilon$ values, where $\epsilon$ and $\delta$ were measured as percentages proportional to the respective range of neural state variables observed in the test data. While computational limits prevent our ability to generate infinitely long trajectories, we used a prediction horizon of length $T=300$, five times longer than videos that make up our datasets, to provide a reasonable substitution. Although our framework cannot be used to definitively identify the stable equilibrium states due to the finite data our models have been trained on, we have found in our experiments that the stable equilibrium states identified through our relaxed approximations were largely consistent with those expected from classical analyses.   
The full equilibrium identification and stability analysis algorithms are given in Section~\ref{appendix:equilibrium_identification} of the \nameref{appendix}.
Next, by calculating all derivatives of the neural state vector field with respect to the state variables at the stable equilibrium state $\vb{V}_{\mathrm{eq}}$, we obtained the Jacobian matrix $\vb{J}$, with which we can linearize Equation~\eqref{eq:vector_field} for states $\vb{V}_t$ near $\vb{V}_{\mathrm{eq}}$, resulting in a linear system
\begin{align}\label{eq:linearized_dynamics}
    \frac{\dd\vb{V}_t}{\dd t} =  \vb{J} \left(\vb{V}_t - \vb{V}_{\mathrm{eq}}\right).
\end{align}
We then estimated the system's natural frequency from the eigenvalues of the Jacobian matrix $\vb{J}$. Details of the experiment parameters and setups are described in Section~\ref{appendix:equilibrium_identification} of the \nameref{appendix}. 

\begin{figure}[H]
    \centering
    \includegraphics[width=1\textwidth]{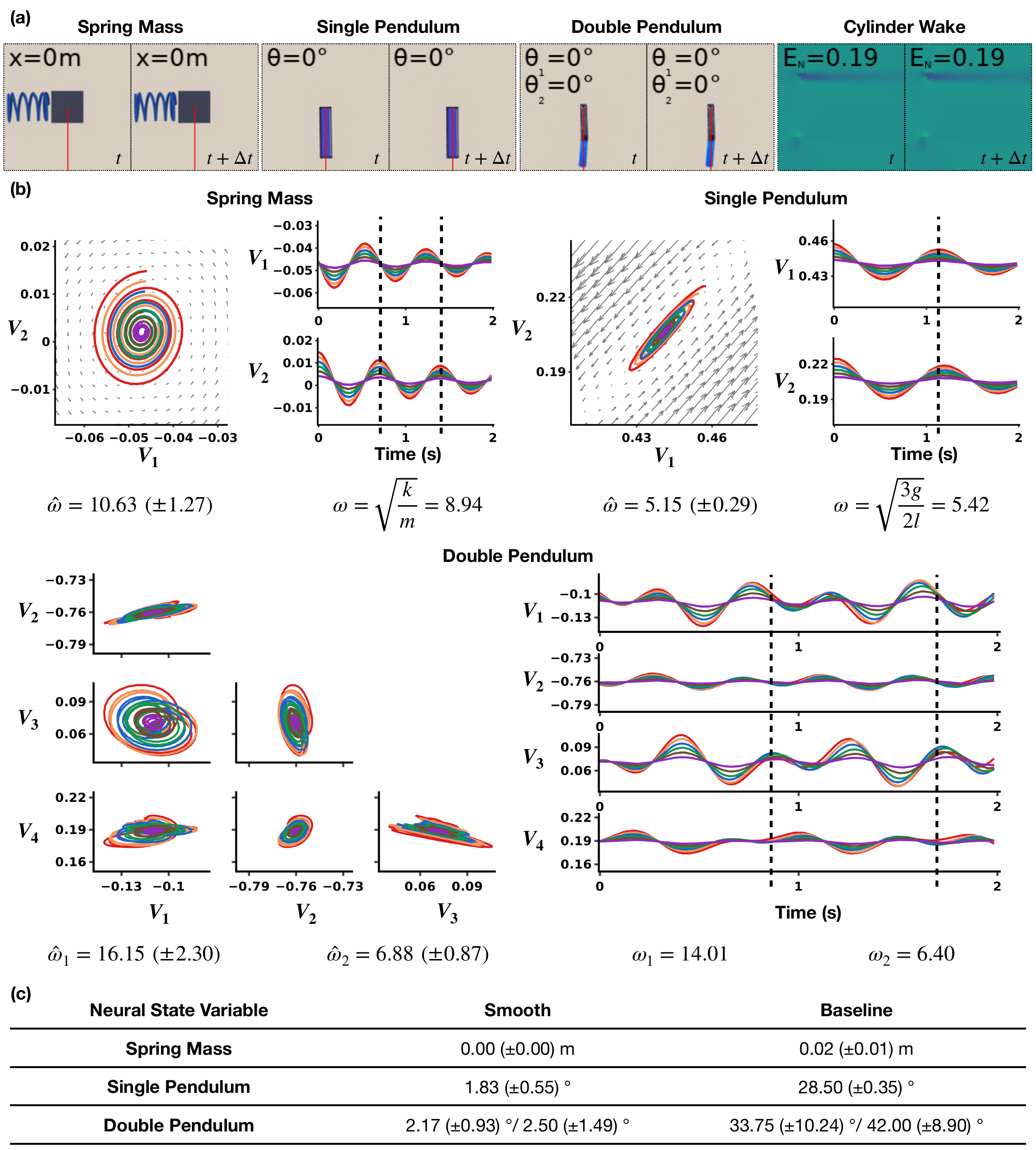}
    ~
    
    \caption{Near-equilibrium analysis. (a) The identified stable equilibrium states are decoded to two frames of images and marked with estimated physical values through computer vision techniques: position for the spring mass, angle for single and double pendulum, and normalized energy for the cylinder wake. Red vertical lines indicate zero positions or angles. (b)  The stability of the equilibrium states is demonstrated for the spring mass, single pendulum and double pendulum systems. 
    We sampled 6 initial states within a distance $\delta$ from the equilibrium ( $\delta=1\%$ for the spring mass and single pendulum, and $\delta=0.5\%$ for the double pendulum) and integrated their neural state vector fields within a region $\epsilon=1.5\%$ centered around the equilibrium state, where $\delta$ and $\epsilon$ are measured in proportion to the range of the neural state variables observed in the test data. 
    The expected natural frequencies $\omega$ and the mean estimated natural frequencies $\hat{\omega}$ along with their respective standard errors are also shown below each system. 
    (c) The mean absolute error of the predicted equilibrium states, along with their standard error bounds, are compared between the results from neural state vector fields trained on smooth and non-smooth neural state variables. 
    }\label{fig:equilibrium}
\end{figure}

\begin{figure}[H]
\centering
\includegraphics[width=\textwidth]{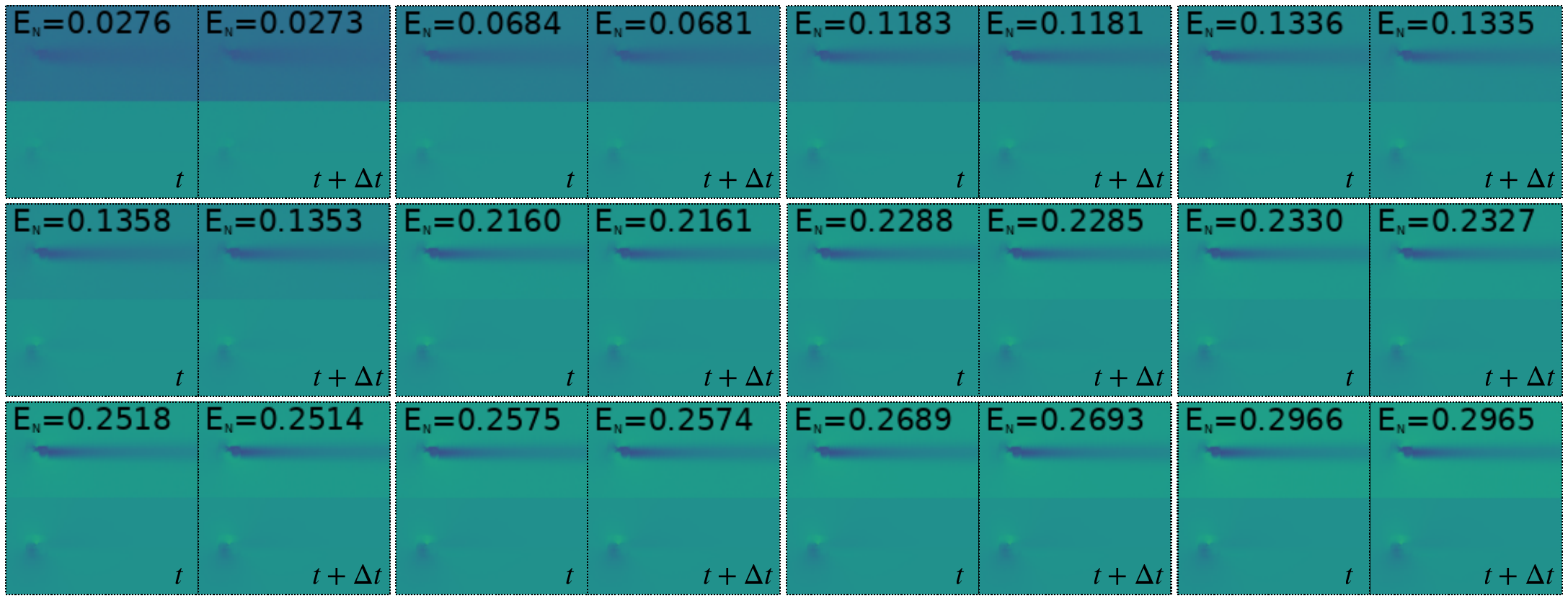}
\caption{Additional identified equilibrium states for the cylinder wake system. More identified stable equilibrium states for the cylinder wake system are shown with their estimated energy, normalized between 0 and 1.}\label{fig:cylindrical_equilibriums}
\end{figure}

Figure~\ref{fig:equilibrium}(a) shows the images decoded from the identified stable equilibrium states for the four systems. Each image is marked with estimated physical quantities from a computer vision algorithm.
For the spring mass system, the estimated position of the mass at the discovered stable equilibrium state is zero. Similarly, for the single pendulum and double pendulum systems, the estimated angles of the arms are zero. Our method accurately identifies the unique stable equilibrium states of these three systems.
The cylinder wake system has non-unique stable equilibrium states, and Figure~\ref{fig:equilibrium}(a) shows one such state successfully identified by our method. More identified equilibrium states are shown in Figure~\ref{fig:cylindrical_equilibriums}. Notably, the discovered equilibrium states may not appear in the training data. Our method both automatically identifies these previously unseen states and renders them into human interpretable video frames. 

Figure~\ref{fig:equilibrium}(b) further demonstrates the stability of the identified stable equilibrium states for the spring mass, single pendulum, and double pendulum systems, through example trajectories plotted in the neural state variable space and with respect to time $t$. For each system, we integrated their respective neural state vector field for $T=120$ (2.0 s), starting from $6$ evenly spaced initial states sampled within a distance $\delta$ to the equilibrium state (e.g. $\delta=1\%$ for the spring mass and the single pendulum, and $\delta=0.5\%$ for the double pendulum). To demonstrate their stability, all plots were cropped to a region $\epsilon=1.5\%$ centered around their respective equilibrium states. We used a longer prediction length compared to that of our dataset (e.g. $T=60$ (1.0 s)) to highlight the long-term stability and to include at least one full period for the trajectories, as marked with a dotted line. Below their respective plots, we also show the ground truth natural frequencies $\omega$ next to our estimated values $\hat{\omega}$ averaged across three random seeds, which are presented along with their respective standard errors. For the spring mass system, the ground truth frequency is $8.944$ and our estimation is $10.629\, (\pm 1.267)$. For the single pendulum system, the ground truth frequency is $5.425$ and our estimation is $5.157\, (\pm 0.290)$. The double pendulum has two normal modes with frequencies $14.007$ and  $6.404$, and our estimations are $16.149\, (\pm 2.299)$ and $6.881\, (\pm 0.871)$ respectively. 
For more details of the calculations for the ground truth frequencies, please refer to Section~\ref{appendix:dataset} of the \nameref{appendix}.

Figure~\ref{fig:equilibrium}(c) compares the mean absolute error of the identified stable equilibrium states from neural state vector fields trained on smooth neural state variables to those trained on baseline neural state variables, which do not incorporate our smoothness constraints. The errors were averaged across three random seeds, and we also provide their respective standard errors. We applied the same equilibrium analysis 
algorithms
with the same parameters for both neural state vector fields. 
The stable equilibrium state is not unique for the cylinder wake system, therefore the error is not reported. The results confirm that enforcing smoothness significantly improves both the accuracy and reliability of the learned neural state vector field for equilibrium identification. The precision of the learned vector field highlights the expressive power of our framework and underscores its potential for conducting quantitative analysis of the near-equilibrium system dynamics. 

Figure~\ref{fig:cylindrical_equilibriums} shows 12 additional examples of stable equilibrium states that were identified by our method for the cylinder wake system. Unlike the mechanical systems, in which there exists a unique globally stable equilibrium state, there exist infinitely many stable equilibrium states in this fluid dynamical system. By simply expanding the number of candidates provided to 
our equilibrium analysis algorithms,
our framework is capable of identifying and visualizing a multitude of such stable configurations. Furthermore, these candidate states are determined without relying on any prior knowledge, and purely through qualitative insights made possible through our method, which are discussed in the next section. More details on how the additional candidate states were chosen are discussed in Section~\ref{appendix:equilibrium_identification} of the \nameref{appendix}.

These results demonstrate that our discovered smooth neural state variables and neural state vector field can describe the underlying dynamics with a remarkable level of detail, without requiring symbolic equations, physical measurements, or human annotations. As a result, our automatically derived operable representations provide accurate analyses of near-equilibrium physics, that not only enable equilibrium discovery from raw visual data but also provide insight into the system's higher-order derivative properties, which are not immediately apparent in the original high-dimensional observations. In the next section, we extend this analysis to more complex dynamics, including chaotic regimes.

\subsection{Non-equilibrium analysis}\label{sec:non-equilibrium}

As a system departs from its stable equilibrium state, its dynamics become increasingly complex, giving rise to richer behaviors.
For example, when the single pendulum is released far from its stable equilibrium state, which is the lowest-energy configuration, the system exhibits periodic motion with a period that increases with amplitude.
For the double pendulum system, its stable equilibrium state is also the lowest-energy configuration. Departing from this equilibrium state, the system energy increases and its dynamics shifts from mild periodic motion to wild chaotic motion. In the latter regime, the system's orbits are no longer periodic or drawn to any equilibrium state or limit cycle; instead, they spread throughout the state space, and small initial perturbations lead to large deviations, making the system’s behavior highly unpredictable.
Likewise, the cylinder wake system undergoes a transition from a laminar steady wake, where the system's orbits converge to stable equilibrium states, to laminar vortex shedding, where the orbits converge to stable limit cycles, as the system's Reynolds number increases.

Analyzing dynamics in the non-equilibrium region is particularly challenging due to coexisting diverse behaviors and potential transitions between them. System dynamics can become highly sensitive to initial conditions, leading to greater sensitivity to model errors when performing the analysis. Furthermore, the lack of reliable visualizations or quantitative metrics makes the analysis far more difficult with raw observations. As a result, it is often impractical and imprecise to detect non-equilibrium behaviors such as limit cycles and chaos from video frames.
Our smooth neural state variable and neural state vector field representations overcome these difficulties by enabling both qualitative and quantitative methods for detecting non-equilibrium behaviors. Using these operable representations, we can analyze non-equilibrium system dynamics by observing such properties as the presence of periodic orbits and limit cycles, and by distinguishing regular and chaotic behaviors through the system orbits' divergence under small perturbations and state space coverage.

\begin{figure}[H]
    \centering
    \includegraphics[width=\textwidth]{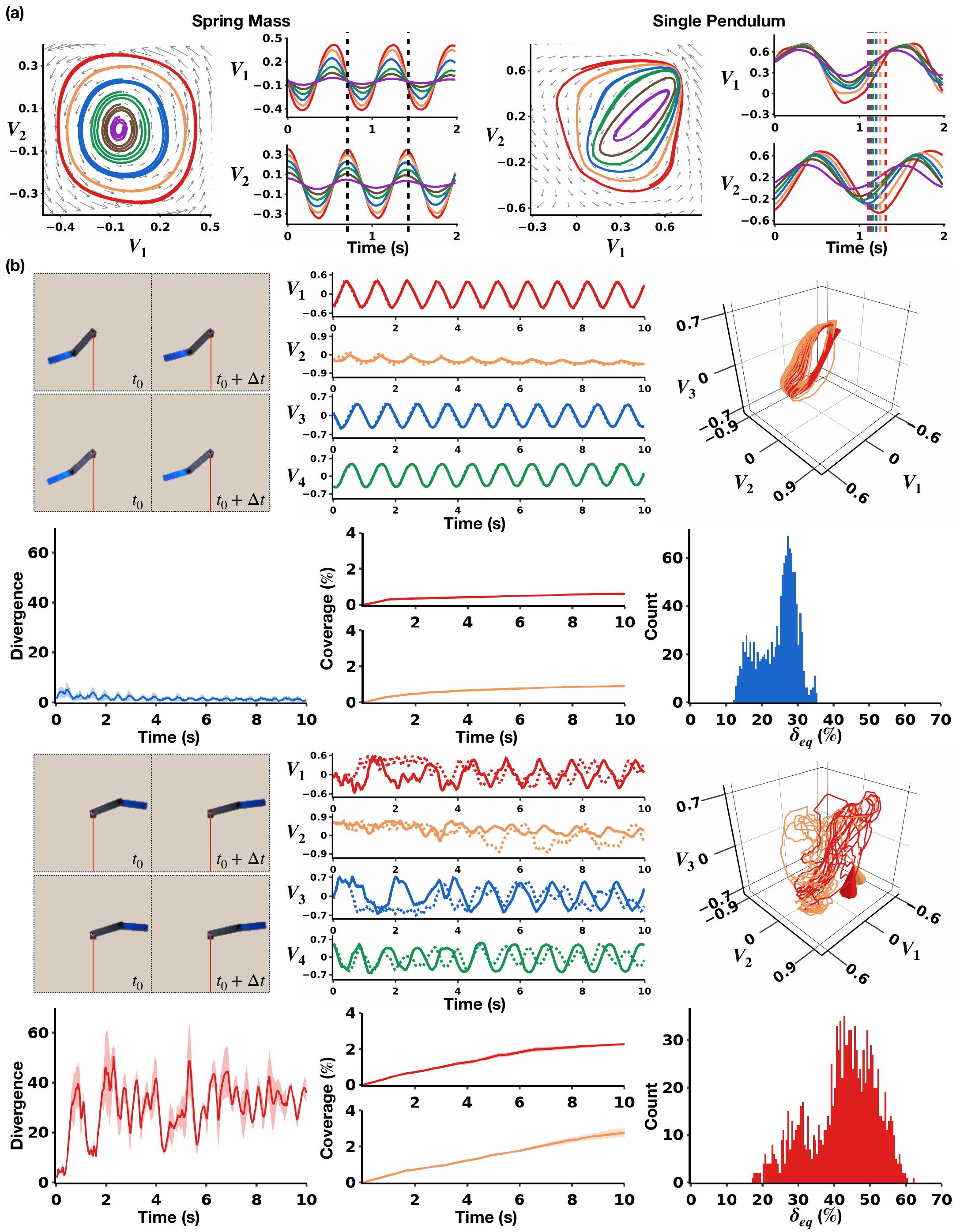}
    \vspace{-6pt}
    \caption{Non-equilibrium analysis. (a) Spring mass and Single pendulum:  We show sampled trajectories, with their approximate periods marked by a dotted line, demonstrating the fixed frequency of the spring mass and varying frequencies of the single pendulum system. (b) Double pendulum: We show the initial states, variables plotted against time, and trajectories plotted by the first three dimensions for pairs of regular (top) and chaotic (bottom) trajectories with nearly identical initial states.  Red vertical lines indicate the zero angles. Below each pair of trajectories, we quantitatively analyze the divergence, state space coverage, and distance to the identified stable equilibrium. Chaotic trajectories have greater sensitivity to initial perturbation, visit more states and remain further from the equilibrium than regular trajectories.}\label{fig:non_equilibrium}
\end{figure}

As our neural state vector field presents an accurate representation of the system dynamics, we can observe integrated trajectories to uncover periodic behavior within the system. In Figure~\ref{fig:non_equilibrium}(a), we display several neural state variable trajectories for the spring mass and single pendulum system, generated by randomly sampling initial states in the smooth neural state variable space and integrating the system using the neural state vector field $\hat{F}$. For both systems, we visualize the trajectories in the smooth neural state variable space and plot each variable, $V_1(t)$ and $V_2(t)$, with respect to time $t$. 
Both systems exhibit potentially periodic behavior, as indicated by the approximately closed loops in the smooth neural state variable space and the repeating patterns in the plots of $V_1(t)$ and $V_2(t)$ over time, both showing that the trajectories return to their initial states after a predictable period. While minor error in our learned model may lead to unclosed loops, which prevent our framework from strongly concluding the presence of periodicity, the trajectories themselves are nearly closed, such that a boundary value problem can be solved to identify periodic orbits up to a reasonable margin of error. More detailed analyses are provided in Supplementary Section~\ref{appendix:non_equilibrium}.
Furthermore, these trajectories uncover characteristic behaviors that can help form the initial analyses of the dynamical system. For the spring mass system, the length of the period is the same for all trajectories, as marked by a black dotted line. A similar periodic pattern can be observed for the single pendulum system.  However, unlike the spring mass system, the length of the period is not consistent for different trajectories. For each trajectory, the approximate length of the period is marked with a dotted line of the same color. From the trajectories, we can observe that the period increases with the amplitude and that the difference in period is more distinct between trajectories with higher amplitudes. This behavior can also be observed in Figure~\ref{fig:equilibrium}(b), where the near-equilibrium trajectories for the single pendulum all display nearly identical period lengths. These qualitative results are consistent with each system's respective classical mechanical analyses, as the spring mass is a linear oscillator with fixed frequency and the single pendulum is a nonlinear oscillator with amplitude-dependent frequencies. 


Beyond periodic orbits, our smooth neural state variables can also help distinguish between regular and potentially chaotic behaviors. Our learned dynamics may not allow rigorous justification for defining chaos; however, the analyses derived from our framework can help discover initial points of interest to further explore the possible presence of chaos within the system. In Figure~\ref{fig:non_equilibrium}(b), we compare two pairs of neural state variable trajectories with nearly identical initial states for the double pendulum system, through visualizations of the encoded variables $V_1(t)$, $V_2(t)$,$V_3(t)$,and $V_4(t)$ with respect to time $t$ and of the trajectories plotted in the $V_1$, $V_2$, and $V_3$ dimensions. The corresponding image frames are shown alongside the plots. 

In the first pair, both trajectories exhibit regular behavior, and their initial deviation remains relatively small throughout the sequence. This periodic behavior is further exemplified as both trajectories remain in a confined region of the state space.

In contrast, the second pair of trajectories drastically diverge over time despite their similarly small initial deviation, thus demonstrating their potentially chaotic behavior and sensitivity to initial perturbation. Furthermore, the trajectories are disordered, spreading over the space and never converging to an equilibrium state or a limit cycle.

To numerically evaluate these observations, we introduced three quantitative metrics: (1) divergence between pairs of trajectories, (2) state space coverage of trajectories, and (3) distance to the stable equilibrium. The divergence metric tracks the Euclidean distance between two trajectories across time, relative to their initial deviation. The coverage metric quantifies how much of the smooth neural state variable space each trajectory explores, calculated by dividing each dimension into $N$ bins (e.g., $N=10$), yielding $N^d$ total boxes, and measuring the fraction visited by the trajectory. Additionally, for each state within a trajectory, we measured its Euclidean distance to the stable equilibrium state identified in \nameref{sec:equilibrium}. Plots depicting the divergence and coverage metrics with respect to time are shown below each respective pair, confirming our qualitative analyses. Next to these plots, we also show a plot depicting the distribution of individual states' distance to the stable equilibrium $\delta$, measured as percentages proportional to the neural state variable space domain. The histograms indicate that regular trajectories remain closer to the stable equilibrium than chaotic trajectories. 

\begin{figure}[H]
    \centering
    \includegraphics[width=\textwidth]{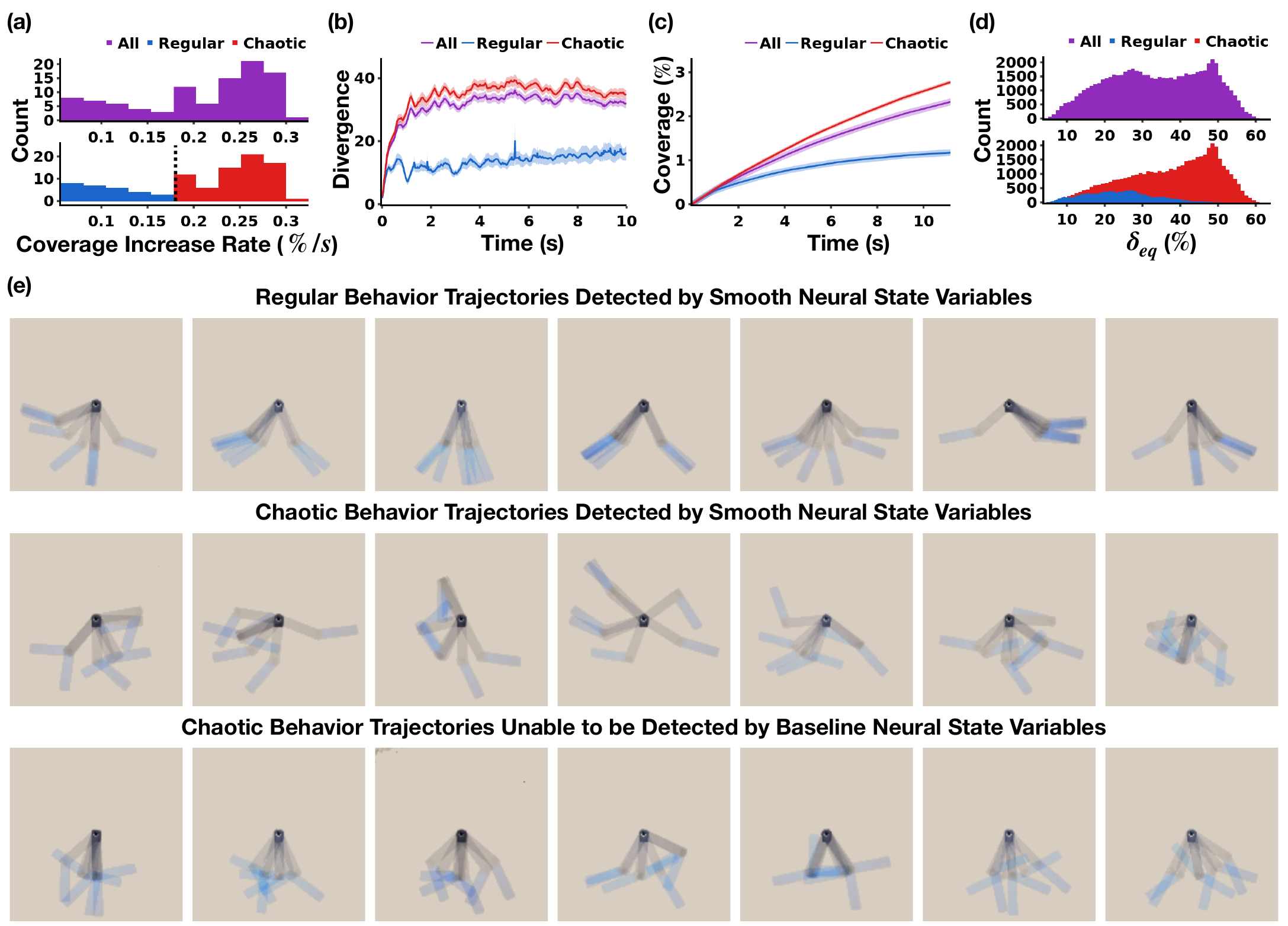}
    \caption{Chaos detection in the double pendulum. (a) A histogram shows the distribution of the mean coverage increase rate of 100 long sequence videos, along with the threshold separating regular and chaotic behavior marked with a dotted line. (b) The mean divergence observed from regular and chaotic trajectory pairs are plotted with respect to time, demonstrating chaotic trajectories' higher sensitivity to initial perturbations. (c) The mean coverage of regular and chaotic trajectories are plotted with respect to time, demonstrating that regular trajectories remain within a more confined region of the state space. (d) A histogram of mean state distance to the identified stable equilibrium demonstrates that chaotic behavior occurs further from the stable equilibrium than regular behavior. (e) Example videos of identified regular and chaotic trajectories, classified using the identified mean coverage increase rate separating threshold, are shown through blended sample frames. Baseline neural state variables are unable to distinguish some chaotic trajectories, as shown in the bottom row.}\label{fig:double_pendulum_chaos}
\end{figure}

We further used the divergence and coverage metrics to detect chaotic behavior in the smooth neural state variable space. Figure~\ref{fig:double_pendulum_chaos} shows the results of computing the metrics for $100$ long-sequence trajectories of the double pendulum system. As shown in the histogram in Figure~\ref{fig:double_pendulum_chaos}(a), the average coverage increase rate over each sequence displays a long-tail distribution, demonstrating the presence of two distinct trajectory behaviors, where some trajectories spread to a broad area in the state space while some remain within a confined region. By applying the k-means algorithm on the average coverage increase rates over each encoded trajectory, we categorized the videos into two groups, one regular and the other chaotic. The identified separating threshold averaged over three random seeds is plotted using a black dotted line. By utilizing these categorizations, we analyzed the encoded trajectories to achieve a broader understanding of the underlying dynamics.

Figure~\ref{fig:double_pendulum_chaos}(b) shows the mean divergence plotted with respect to time, where the divergence was averaged over all pairs of trajectories with initial states closer than $1\%$ of the neural state variable range observed in the test set. The mean divergence averaged over pairs of trajectories separated by category are plotted as well. The plot indicates that the identified chaotic trajectories show greater sensitivity to initial perturbations, validating the categorizations. Moreover, the mean coverage plotted with respect to time in Figure~\ref{fig:double_pendulum_chaos}(c) shows a stark decrease in slope only for the identified regular trajectories, indicating that regular trajectories remain within a confined region of the state space by reaching a limit cycle. On the other hand, chaotic trajectories do not converge to any equilibrium states or limit cycles, and they keep exploring a broader region of the state space. Finally, in Figure~\ref{fig:double_pendulum_chaos}(d), the histogram plotting the mean state distance to the stable equilibrium 
demonstrates that chaotic trajectories contain a high concentration of states further from the stable equilibrium, whereas states in regular trajectories are concentrated in a region that is relatively closer to the stable equilibrium. These results are consistent with the analysis from classical mechanics, suggesting that while the double pendulum system exhibits stability and predictability in low-energy configurations, it becomes increasingly sensitive to initial conditions and displays chaotic behavior as the system's energy increases. Figure~\ref{fig:double_pendulum_chaos}(e) visualizes example videos for the detected regular and chaotic behavior through time-elapsed frames. The final row demonstrates the baseline model's inability to detect some chaotic behavior when the same methods are applied. 


\begin{figure}[H]
    \centering
    \includegraphics[width=\textwidth]{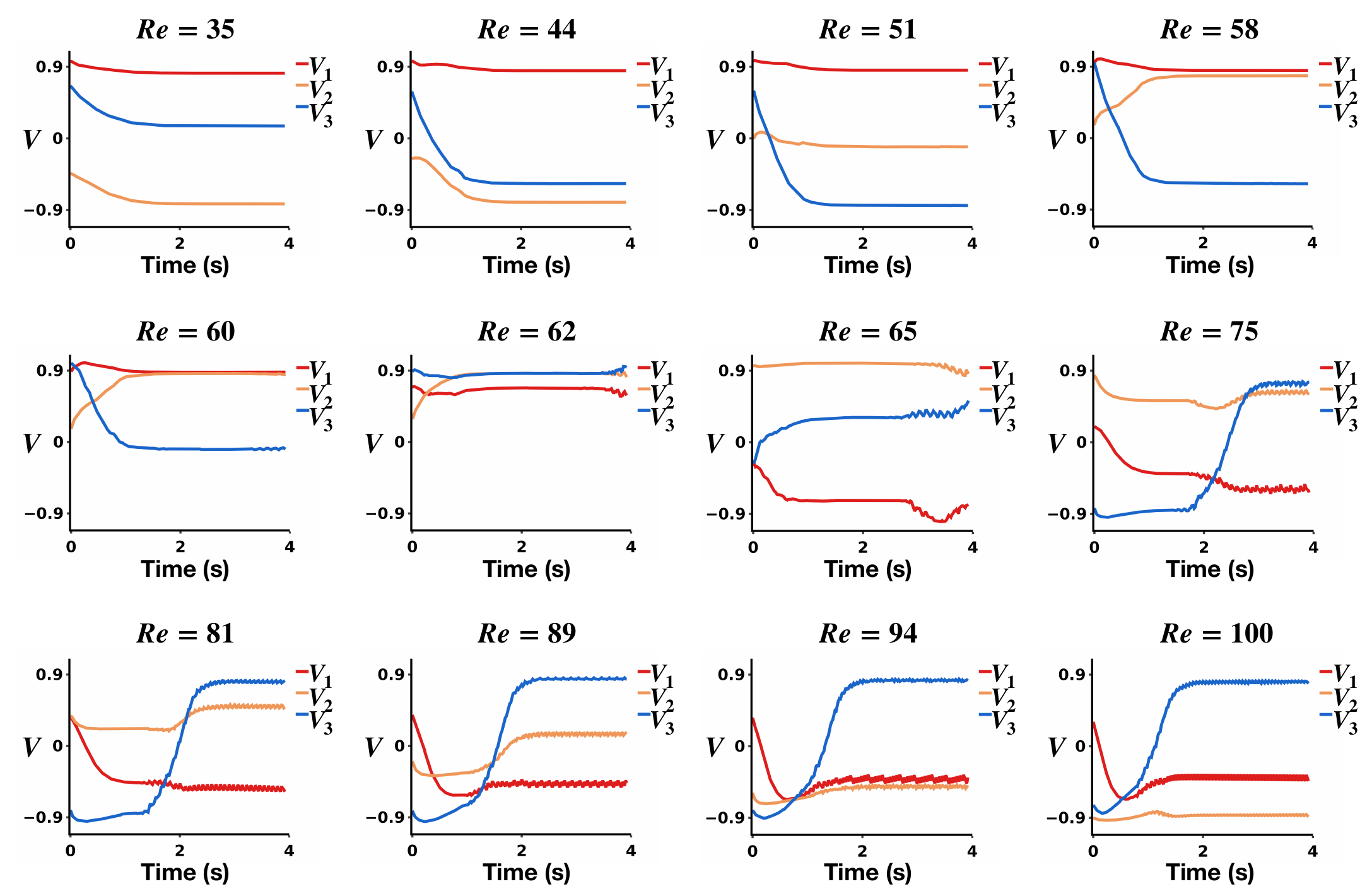}
    \caption{System bifurcation in the cylinder wake. We show plots of encoded trajectories with increasing Reynolds numbers for the cylinder wake system, demonstrating the system bifurcation from laminar steady wake to laminar periodic vortex shedding behaviors as the Reynolds numbers increase.}\label{fig:cylindrical_non-equilibrium}
\end{figure}

The flexibility of our approach allows our smooth neural state variables to not be limited to mechanical systems, and they can also provide meaningful analysis of non-equilibrium dynamics for fluid dynamical systems, as demonstrated by the identification of the cylinder wake system's bifurcation to the vortex shedding regime. In Figure~\ref{fig:cylindrical_non-equilibrium}, we present various plots for the neural state variables $V_1(t)$, $V_2(t)$, and $V_3(t)$ for the cylinder wake system, plotted against time $t$. These plots correspond to trajectories from the test data with increasing Reynolds number of the flow. In the first row, with the lowest Reynolds numbers, we observed that the trajectory converges to a stable equilibrium state, indicating a laminar steady wake. In the second and third rows, with increased Reynolds numbers, the trajectories exhibit two phases: initially converging towards a state, and then circulating around it, indicating the presence of a stable limit cycle. In this regime, the trajectory is attracted to the limit cycle from its initial state and then transitions into periodic motion. As the Reynolds number increases, the system converges to a limit cycle more quickly. The change in the system's behavior suggests the presence of a bifurcation from a laminar steady wake to laminar periodic vortex shedding. For more details on the laminar steady wake and laminar periodic vortex shedding regimes of the cylinder wake system, please refer to Section~\ref{appendix:dataset_cylindrical_flow} of the \nameref{appendix}.

By utilizing these qualitative observations from the encoded smooth neural state variable trajectories, our neural state vector field is capable of revealing further insight into the global state space dynamics of the cylinder wake system. As shown in the first row of Figure~\ref{fig:cylindrical_non-equilibrium}, trajectories corresponding to low Reynolds numbers demonstrate that the system converges to a stable equilibrium state. Through applying our near-equilibrium analysis algorithms on these identified states, our framework further confirmed the presence of multiple stable equilibrium states at various energy levels, which are visualized in Figure~\ref{fig:cylindrical_equilibriums}. These analyses of the cylinder wake system highlight our framework's ability to generate a robust representation space that accurately captures the dynamics and global landscape of the system.

\subsection{Synthesizing new data with parameterized  novel dynamics}

\begin{figure}[H]
    \centering
    \includegraphics[width=1\textwidth]{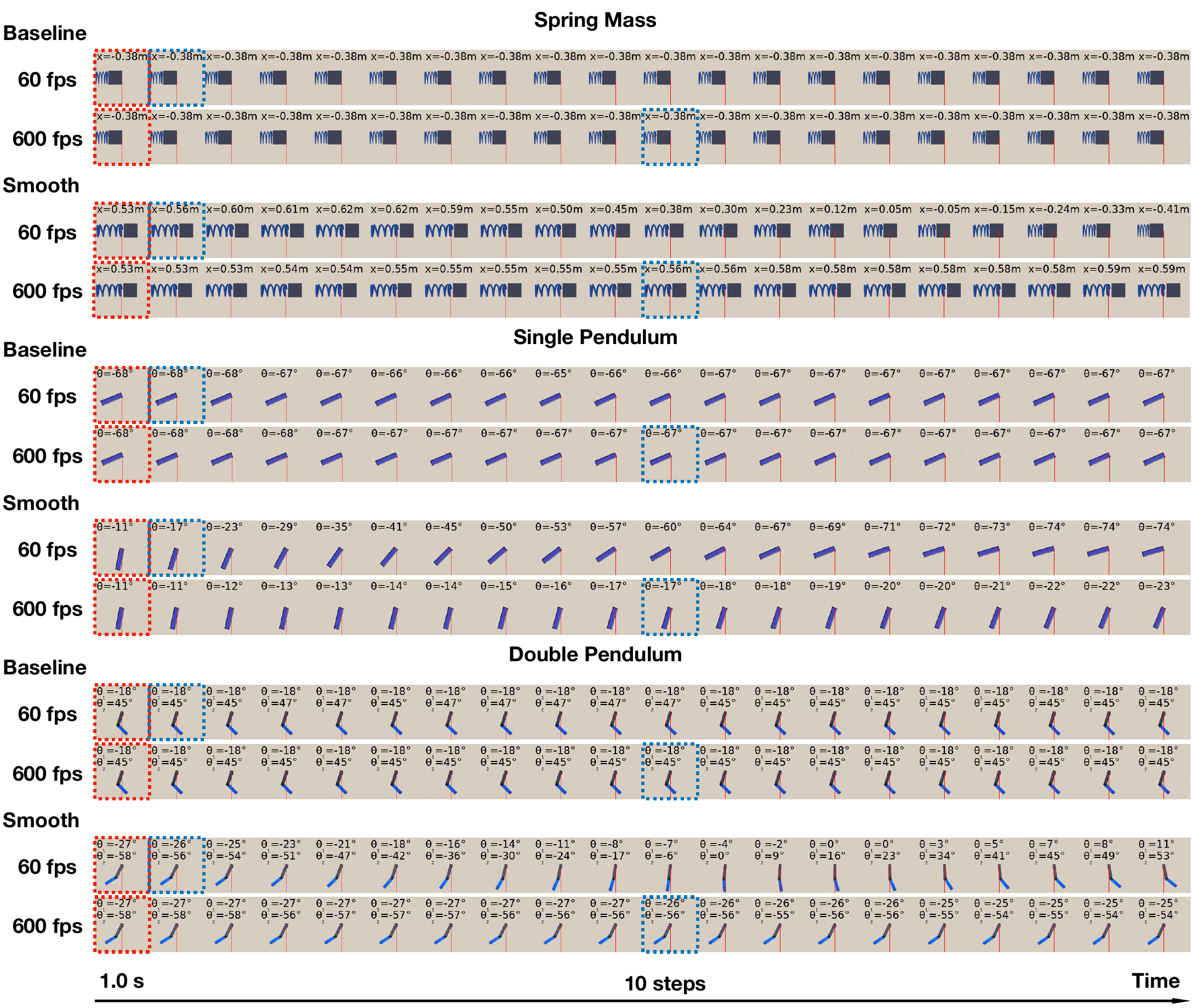}
    ~
    \caption{ Sample decoded video frames from integrated trajectories with varying frame rates. 20 consecutive frames in varying rates are decoded from neural state vector field integrated trajectories with randomly sampled initial states for the spring mass, single pendulum and double pendulum systems. Red vertical lines indicate zero positions or angles. To highlight the long term prediction stability, the frames begin from the middle of the videos (~$t=1.0$~s~). 
    The corresponding frames between the different frame rate videos are marked with dotted boxes of the same color. Only smooth neural state variables allow consistently realistic video generation with variable frame rates for all three systems.
    }\label{fig:timeStepVariation}
\end{figure}

In this section, we will further demonstrate the robustness of our derived operable representations. In particular, our smooth neural state variables and neural state vector field are capable of synthesizing stable long term predictions that can not only simulate the dynamics of the original system, but also generate novel behaviors that remain physically plausible, in a well-defined and controllable manner. For example, by leveraging the integrable neural state vector field $\hat{F}$, we can produce new video sequences from randomly sampled initial states with arbitrary frame rates, which are not limited to that of the dataset. Figure~\ref{fig:timeStepVariation} shows sequences of video frames decoded from neural state vector field integrated trajectories with varying frame rates. For smooth neural state variables, the trajectories decoded to physically plausible videos, which remained consistent to the original videos (60~fps) even with an increased frame rate (600~fps). In contrast, baseline neural state variables' trajectories often decode to physically implausible and inconsistent videos.

Additionally, we can generate novel behaviors by leveraging the system's unique stable equilibrium state, which were identified through our near-equilibrium analyses for the spring mass, single pendulum, and double pendulum systems. By introducing an artificial damping term, we can direct the system trajectories towards the stable equilibrium state $V_{\mathrm{eq}}$. Specifically, we construct the following equation:
\begin{align}\label{eq:damped_dynamics}
    \frac{\dd \vb{V}_t}{\dd t} = \hat{F}_\gamma(\vb{V}_t) \doteq \hat{F}(\vb{V}_t) - \gamma (\vb{V}_t-\vb{V}_{\mathrm{eq}}),
\end{align}
where $\hat{F}$ is the neural state vector field of the system and $\gamma\geq0$ is a damping factor. The term $- \gamma (\vb{V}_t-\vb{V}_{\mathrm{eq}})$ acts as an external force, dragging the system trajectories in the smooth neural state variable space toward the equilibrium state $\vb{V}_{\mathrm{eq}}$.

\begin{figure}[H]
    \centering
    \includegraphics[width=1\textwidth]{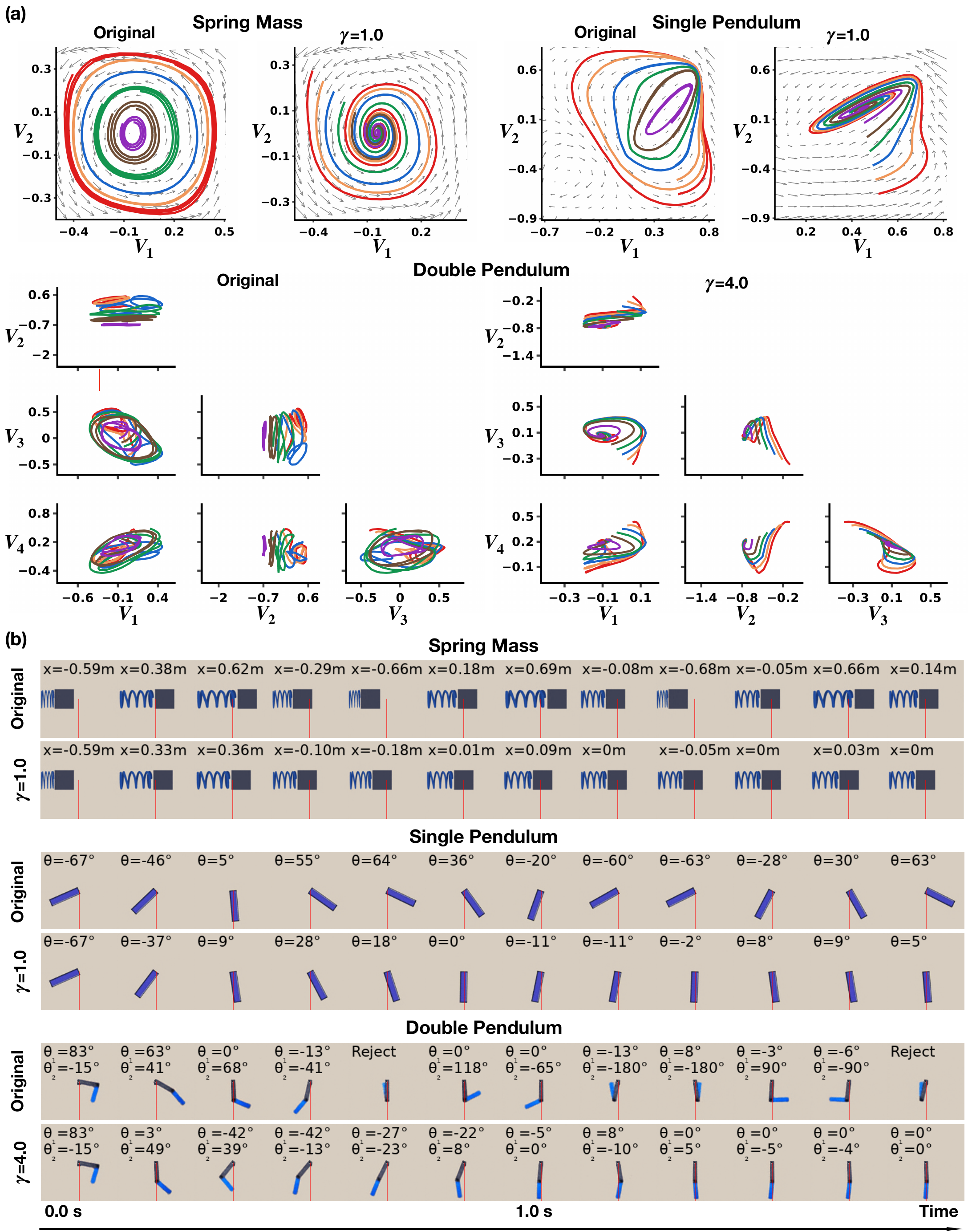} \\
    \caption{Synthesizing physically plausible data sequences with unseen dynamics. (a) Sampled trajectories are plotted in the original and damped smooth neural state variable space for the spring mass, single pendulum and double pendulum systems. (b) Video frames decoded from one of the trajectories exemplify the physically plausible effects of damping for each of the three systems.  Red vertical lines indicate zero positions or angles.}\label{fig:damping}
\end{figure}

A comparison between the original and dissipative dynamics are presented in Figure~\ref{fig:damping} for the spring mass, single pendulum and double pendulum systems. Figure~\ref{fig:damping}(a) shows example trajectories plotted in the smooth neural state variable space for each system. These trajectories were generated by randomly sampling initial states and integrating their respective dynamics, as defined by Equation~\eqref{eq:vector_field} and Equation~\eqref{eq:damped_dynamics}, with a non-zero damping factor $\gamma$ (e.g. $\gamma = 1.0$ for single pendulum and spring mass, $\gamma = 4.0$ for the double pendulum). For the spring mass and single pendulum systems, the gradient fields for the original and perturbed dynamics are also plotted with the trajectories. The plots reveal that the new system trajectories display damped patterns and are attracted to the equilibrium state when $\gamma > 0$. In Figure~\ref{fig:damping}(b), we present video frames generated from one of the trajectories for each of the spring mass, single pendulum and double pendulum systems shown in Figure~\ref{fig:damping}(a), sampled at 6 frames per second (fps). The video frames show that the damped dynamics attract the systems towards their stable equilibrium states. 

Our results demonstrate the potential of our method to synthesize new, physically plausible behaviors, including dynamics not observed in the training data. Furthermore, the generated novel physics remains interpretable, and the effects of dynamics modifications are fully adjustable through the damping parameter $\gamma$. More results for novel physics generation with varying $\gamma$ are shown in Section~\ref{appendix:new_data_generation} of the \nameref{appendix}.
Although synthesizing such new dynamics is straightforward with known physical variables and equations, our approach achieves this directly from raw video data by utilizing the learned well-structured physically meaningful state space. This capability underscores the robustness of our derived representations as trajectories that deviate from those seen in the original data can still decode to physically plausible videos, indicating the accurate global landscape of the learned state space. 

\section{Discussion}\label{sec:discussion}

We have developed a framework capable of extracting and analyzing the dynamics of physical systems directly from high-dimensional observational data captured in videos, without relying on human annotated measurements, symbolic priors, or domain-specific knowledge. Our approach discovers smooth neural state variables and constructs a neural state vector field that together form a compact yet highly expressive representation of system dynamics. Crucially, these representations are not only suitable for prediction, but also operable. They provide an operational basis for applying various mathematical techniques to study the system’s behavior, thus bridging modern data-driven machine learning approaches with the traditionally successful scientific discovery paradigm established among human scientists. 

To extract such representations, our method applies non-intrusive constraints that do not require any prior physics knowledge and rely only on the minimal assumption that the system evolves continuously in time. This temporal continuity is enforced through smoothness constraints that guide the unsupervised training process to automatically recover state variables that accurately reflect the system's true underlying dynamics. Despite their simplicity, our proposed smoothness constraints yield representations rich enough to support derivative-based analysis and reveal higher-order properties that are not immediately apparent in the original high-dimensional observations. 

As a result, our framework allows many classical and powerful techniques to be applied directly on raw high-dimensional observational data for scientific discovery. It enables the automated discovery of stable equilibrium states, linearized dynamics around these equilibrium states, and detection of key system characteristics such as natural frequencies, 
as well as potentially periodic, limit cycle, and chaotic behaviors in the non-equilibrium regime. Additionally, our robust representation space supports the synthesis of physically plausible data with previously unseen dynamic behaviors in the form of videos, by perturbing the system dynamics. These capabilities highlight a critical shift in how AI systems can contribute to scientific discovery, not by relying on hand-crafted priors or hard-coded physical constraints, but by autonomously uncovering compact, operable representations directly from raw observations. In doing so, our approach augments the prevailing use of  physics-informed architectures or symbolic regression by offering a powerful complementary path for discovery that begins with minimal assumptions.

However, while our proposed framework aims to accelerate scientific discovery of dynamical systems without system-specific assumptions, the fidelity of the reconstructed trajectories, along with any insights into potential system properties, depend on two core working assumptions: (i) the system is structurally stable, i.e., its qualitative long-term behavior remains unchanged under small perturbations to the governing equations (see ref.~\cite{smale1966structurally} and references therein for the limited prevalence of structurally stable systems, particularly in high dimensions); (ii) the video data provide adequate spatiotemporal coverage and temporal resolution. We acknowledge that these assumptions necessitate rigorous analytical justification. Consequently, analyses of long-term dynamics remain suggestive rather than definitive absent further analytical verification, also being confined to available training data. Nevertheless, we believe our framework will serve as a powerful tool for automating the initial stages of scientific discovery, laying a foundation for deeper rigorous analysis.

Future directions for this approach are vast. One promising avenue is to apply our method to more complex systems across diverse fields, such as turbulent fluid dynamics and quantum mechanics in physics, reaction kinetics and combustion in chemistry, or cellular and neural processes in biology. Another important direction is extending our framework to handle systems with sudden changes, such as collision systems. Integrating human expertise into this computational loop could further accelerate discovery through collaboration between domain experts and AI. Moreover, the discovered smooth neural state variables could be useful for controlling these complex systems. Their interpretability, along with their ability to be extracted directly from high-dimensional raw observations, presents a compelling middle ground between black-box learning and model-based control. 

Another important direction for future research lies in further enhancing the interpretability of the discovered smooth neural state variables and neural state vector field. By combining our approach with symbolic regression techniques, we could extract concise, human-readable equations that can illuminate the underlying principles governing the observed phenomena. This refinement could bridge theoretical gaps in fields with abundant data but incomplete models, such as biology and cosmology. To this end, our work provides a solid foundation for aiding human scientists with the power of AI and accelerating scientific progress.

\section{Methods}\label{sec:methods}

In this section, we detail our methods for extracting neural state variables from high-dimensional image data while preserving the smoothness of system dynamics and for constructing a neural state vector field to describe the system dynamics. Further implementation details, including a full description of model architectures used, are provided in Section~\ref{appendix:implementation_details} of the \nameref{appendix}.

\subsection{Extracting neural state variables}\label{sec:method:variables}
To automatically derive smooth neural state variables of the most compact form, we trained two auto-encoder neural networks $g=(g_E,g_D)$ and $h=(h_E,h_D)$. The first network $g=(g_E,g_D)$ encodes two consecutive frames $\mathbf{C}_t = (\mathbf{X}_t, \mathbf{X}_{t+\Delta t})$ to a latent vector $\mathbf{L}_t$ and decodes the latent vector to predict the next two consecutive frames $\mathbf{C}_{t+2\Delta t} = (\mathbf{X}_{t+2\Delta t}, \mathbf{X}_{t+3\Delta t})$. The auto-encoder $g$ was trained by minimizing the pixel reconstruction loss between the ground truth $\mathbf{C}_{t+2\Delta t} = (\mathbf{X}_{t+2\Delta t}, \mathbf{X}_{t+3\Delta t})$ and the predicted output $\hat{\mathbf{C}}_{t+2\Delta t} = (\hat{\mathbf{X}}_{t+2\Delta t}, \hat{\mathbf{X}}_{t+3\Delta t})$ from $g$. We then estimated the system's intrinsic dimension $d$, which is the minimum number of state variables necessary for describing the system dynamics \cite{chen_automated_2022}, by applying the Levina-Bickel's algorithm \cite{levina2004maximum} to the latent vectors $\mathbf{L}_t \in \mathbb{R}^N$ encoded from the test set. To increase the confidence of the derived intrinsic dimension, we applied the dimension analysis to two different values of the bottleneck dimension $N$, which were chosen to be 64 or 8192. For the estimated intrinsic dimensions of the four studied systems, please refer to Section~\ref{appendix:intrinsic_dim} of the \nameref{appendix}.
where the estimated intrinsic dimension values are rounded to their nearest integers. Next, we trained a second auto-encoder $h=(h_E,h_D)$ that further compresses the latent vectors $\vb{L}_t$ to produce neural state variables $\vb{V}_t \in \mathbb{R}^d$, where $d$ is the system's intrinsic dimension. We utilized the latent vectors with $N = 64$, for this compression step. We discuss the training of $h$ in further detail in the following subsection. The model architectures for auto-encoders $g$ and $h$ are outlined in Section~\ref{appendix:model_architecture_g_and_h} of the \nameref{appendix}.

\subsection{Enforcing smoothness on neural state variables}\label{sec:method:smoothness}

The auto-encoder $h=(h_E,h_D)$ encodes a latent vector $\vb{L}_t$ to its corresponding neural state variable $\vb{V}_t$ and decodes $\vb{V}_t$ to reconstruct the input $\vb{L}_t$.
The training objective is not only to minimize the reconstruction loss  on the latent vector $\mathbf{L}_t$, but also to ensure the resulting state variables $\vb{V}_t$ are operable representations such that human scientists can conduct further analyses by applying mathematical tools such as calculus on the extracted state space. While many methods have been explored for achieving such interpretable embeddings \cite{champion2019data,conti2024venivindyvicivariational,bakarji2022discoveringgoverningequationspartial,Conti_2023,greydanus2019hamiltonian,cranmer_lagrangian_2020}, in order to fully preserve the underlying dynamical information of the system without imposing prior assumptions that may otherwise distort or influence the derived dynamics, we must only rely on minimal constraints based on first principles, such as continuity with respect to time. To achieve this goal, we propose to enforce smoothness of neural state variable trajectories $\{\vb{V}_0,\vb{V}_{\Delta t},\vb{V}_{2\Delta t},\cdots\}$ by minimizing distances between neighboring states on these trajectories. Concurrently, we must also prevent the smoothness constraint from dominating the training objective, which may cause all neural state variables to collapse to a single point.
In this extreme case, distances between neighboring states on all trajectories become zero and the system dynamics information is totally lost.

Given these considerations, we minimized a loss function $\mathcal{L}$ that is composed of three loss terms: a reconstruction loss, a smoothness loss, and a space-filling loss:
\begin{align}\label{loss_func}
    \mathcal{L} = w_{\mathrm{reconstruct}}\mathcal{L}_{\mathrm{reconstruct}} + \beta \left( w_{\mathrm{smooth}} \mathcal{L}_{\mathrm{smooth}} + w_{\mathrm{space}} \mathcal{L}_{\mathrm{space}} \right).
\end{align}
The reconstruction loss $\mathcal{L}_{\mathrm{reconstruct}}$ is the sum of squared errors between the latent vector $\mathbf{L}_t$ and the reconstructed $\hat{\mathbf{L}}_t$ from the auto-encoder $h$.
The smoothness loss $\mathcal{L}_{\mathrm{smooth}}$ minimizes distances between neighboring states on neural state variable trajectories.
Formally, we define:
\begin{align}
    \mathcal{L}_{\mathrm{smooth}}&= \max\left(0, \ \norm{\vb{V}_{t+2\Delta t} - \vb{V}_t} - 2L_0 \right) + \eta \cdot \max\left(0, \ \norm{\vb{V}_{t+\Delta t} - \vb{V}_t} - L_0 \right), \label{smooth_loss_2}
\end{align}
with a choice of only penalizing the distance between states $\vb{V}_t$ and $\vb{V}_{t+2\Delta t}$ when $\eta=0$ or the distances between states $\vb{V}_t$, $\vb{V}_{t+\Delta t}$ and $\vb{V}_{t+2\Delta t}$ when $\eta=1$, where $L_0>0$ is a threshold parameter and $\norm{\cdot}$ denotes the Euclidean distance in $\mathbb{R}^d$. To help prevent the collapse of the encoded trajectories, we only enforced that those distances between neighboring states be less than given threshold values.
The use of such threshold values for enforcing smoothness constraints has also been introduced in ref.~\cite{rosca_case_2021}. However, without further intervention, the smoothness loss alone may induce the collapse of all states into a small region in the state space.
The space-filling loss $\mathcal{L}_{\mathrm{space}}$ is the deviation between the distribution of all encoded data points and the uniform distribution on the domain $[-1,1]^d$ within the neural state variable space, measured by the Sinkhorn distance \cite{feydy2018interpolating}. 
The design of such a loss aims to further prevent neural state variable trajectories from collapsing to a single point by utilizing the fact that the outputs of the sub-layers of the auto-encoder $h$ are restricted to the domain $[-1,1]^d$ due to our use of sinusoidal activation functions. Additionally, the generalizable form of Equation~\eqref{smooth_loss_2} can capitalize on the bounded nature of the embedded variables to extend our discovered state variables to non-Euclidean spaces. By enabling the distance metric to treat the fixed neural state variable domain either as a bounded box or as a continuous torus with periodic boundary conditions, our smoothness constraints provide flexibility in handling states near the domain boundaries.

Finally, combining multiple losses can lead to potential optimization problems, where the training may get stuck in local minima. This problem is similar to the KL vanishing problem when training a variational auto-encoder, and a possible solution is to use an annealing schedule \cite{fu_cyclical_2019, bowman_generating_2016}. 
In our method, we introduced a cyclic annealing schedule for the parameter $\beta$ in equation~\eqref{loss_func} to optimize our multiple losses. During training, we first set the parameter $\beta=0$ 
and trained the auto-encoder $h$ with only the reconstruction loss $\mathcal{L}_{\mathrm{reconstruct}}$. We then gradually increased $\beta$ from zero to one to encourage the optimization of $\mathcal{L}_{\mathrm{smooth}}$ and $\mathcal{L}_{\mathrm{space}}$, and the value $\beta=1$ was fixed for a predefined duration. At the end of this stage, we set $\beta$ back to zero to start another cycle. By cyclically varying the value of $\beta$, the optimization process can better escape local minima when enforcing the smoothness and space-filling constraints.
For more implementation details on the smoothness regularization constraints, please refer to Section~\ref{appendix:enforcing_smoothness} of the \nameref{appendix}.

\subsection{Constructing a neural state vector field}\label{sec:method:vectorfield}

We used a neural state vector field $\hat{F}$, implemented as a multi-layer perceptron, to describe the system dynamics of the discovered smooth neural state variables, as shown in Equation~\eqref{eq:vector_field}. The full model architecture is described in Section~\ref{appendix:learning_nsvf} of the \nameref{appendix}.

Before the training of $\hat{F}$, we first encoded all sequences into smooth neural state variable trajectories. We then filtered out trajectories from each of the training, test, and validation subsets, removing those with any $\norm{\vb{V}_{t+\Delta t} - \vb{V}_t}$ that exceeded their corresponding thresholds, which were set based on the distribution of $\norm{\vb{V}_{t+\Delta t} - \vb{V}_t}$ across their respective subsets, and keeping only those within the 99th percentile. For more details on the effects of this filtering process, please refer to Section~\ref{appendix:learning_nsvf} of the \nameref{appendix}.

Next, we trained $\hat{F}$ by integrating Equation~\eqref{eq:vector_field} and comparing the resulting integrated trajectories with the ground truth encoded trajectories. This was implemented using a NeuralODE \cite{chen2019neuralordinarydifferentialequations, politorchdyn}, aiming to minimize the prediction loss defined by the following loss function:
\begin{align}
    \mathcal{L} = \frac1M\sum_{k=1}^M \sum_{m=0}^{N-2} \frac{\sum_{n=m+1}^{N-1} \rho^{n-m-1}\norm{\hat{\vb{V}}^{(k)}_{m\Delta t\to n\Delta t} - \vb{V}^{(k)}_{n \Delta t}}}{\sum_{n=m+1}^{N-1}\rho^{n-m-1}}, \label{mlp_loss}
\end{align}
where $M$ is the number of remaining smooth neural state variable trajectories after filtering, $N$ is the length of these trajectories, and $\hat{\vb{V}}^{(k)}_{m \Delta t\to n \Delta t}$ is the solution to the initial value problem defined by Equation~\eqref{eq:vector_field}, evaluated at time $n \Delta t$, given the initial state $\vb{V}^{(k)}_{m\Delta t}$ at time $m \Delta t$. For each $m=0,\cdots,N-2$, we minimized the weighted sum of the losses between the predicted state $\hat{\vb{V}}^{(k)}_{m\Delta t\to n\Delta t}$ and the ground truth $\vb{V}^{(k)}_{n \Delta t}$ for $n=m+1,\cdots,N-1$. The weight parameter $\rho<1$ was introduced to prioritize short-term predictions over long-term ones, and $\rho$ was cyclically annealed from $\rho=0.1$ to $\rho=0.9$, similar to the annealing of $\beta$ in Equation~\eqref{loss_func}. 

Alternatively, we could train $\hat{F}$ by directly minimizing the reconstruction loss between $\hat{F}\left(\vb{V}^{(k)}_{n\Delta t}\right)$ and the finite difference $\frac{\vb{V}^{(k)}_{(n+1)\Delta t}-\vb{V}^{(k)}_{n\Delta t}}{\Delta t}$ that approximates the derivative $\frac{\dd\vb{V}_t}{\dd t}$ in Equation~\eqref{eq:vector_field} evaluated at $\vb{V}^{(k)}_{n\Delta t}$. While this approach achieved similar accuracy for short-term predictions, the integration-based approach using the loss function \eqref{mlp_loss} resulted in significantly more accurate long-term predictions. For more details on the improved accuracy, please refer to Section~\ref{appendix:learning_nsvf} of the \nameref{appendix}.

\newpage

\section*{Acknowledgments}\label{sec:acknowledgment}

This work is supported by ARL STRONG program under awards W911NF2320182 and W911NF2220113, by ARO W911NF2410405, by DARPA FoundSci program under award HR00112490372, and DARPA TIAMAT program under award HR00112490419.

\subsection*{Author Contributions}\label{sec:author_contribution}

K.H. and D.C. contributed equally to the manuscript. K.H., D.C., and B.C. conceived and designed the experiments. K.H., D.C., and B.C. collected the data. K.H. and D.C. analyzed the data. K.H. and D.C. contributed to materials and analysis tools. K.H., D.C., and B.C. prepared the manuscript.

\subsection*{Data Availability}\label{sec:data_availability}

All datasets used in this manuscript are available at

\url{https://github.com/generalroboticslab/automated_discovery_of_continuous_dynamics_from_videos}.

\subsection*{Code Availability}\label{sec:code_availability}

All codes for generating the results in this manuscript are available at 

\url{https://github.com/generalroboticslab/automated_discovery_of_continuous_dynamics_from_videos}.

\subsection*{Competing Interests}\label{sec:competing_interests}

The authors declare no competing interests.

\newpage

\begin{appendices}\label{appendix}

\clearpage


\section*{Supplementary Information}\label{appendix}

\renewcommand\figurename{Supplementary Figure}
\renewcommand\tablename{Supplementary Table}%

\renewcommand{\theHfigure}{\arabic{figure}}
\renewcommand{\thefigure}{\arabic{figure}}
\renewcommand{\thetable}{\arabic{table}}
\renewcommand{\theequation}{\arabic{equation}}

\section{Details on datasets}\label{appendix:dataset}
In this section, we provide more information on the physics governing equations and the data collection processes for the four systems under study.
\subsection{Spring mass}\label{appendix:dataset_spring_mass}
The spring mass system describes the motion of an object connected to a spring, and it is governed by the following equations of motion:
\begin{align}
    \dot{x} &= v, \\
    \dot{v} &= -\frac{k}{m}x,
\end{align}
where $x$ is the object's position and $v$ is its respective velocity.
The object's mass is $m=1\si{\kilogram}$ and the spring constant is $k=80\si{\newton/\meter}$. The two variables $x$ and $v$ specify the system's state and form a set of state variables of the system.

The spring mass system is a linear oscillator whose solution is given by:
\begin{align}
    x(t) &= x_0 \cos(\omega t) + \frac{v_0}{\omega} \sin(\omega t), \\
    v(t) &= -\omega x_0 \sin(\omega t) + v_0 \cos(\omega t),
\end{align}
where $\omega=\sqrt{\frac{k}{m}}$ is the natural frequency of the system, and $(x_0,v_0)$ is the system's initial state.
The system has only one equilibrium state at $x=0,v=0$ that is stable.

To collect the data, we randomly sampled the object's initial position and velocity. In total, we collected 1,200 sequences with 60fps. We used 960 of these sequences for training, 120 of them for validation, and 120 of them for testing.

\subsection{Single pendulum}\label{appendix:dataset_single_pendulum}
The single pendulum system is governed by the following equations of motion:
\begin{align}
    \dot{\theta} &= \Omega, \\
    \dot{\Omega} &= -\frac{3g}{2L}\sin\theta,
\end{align}
where $\theta$ is the angle of the pendulum arm and $\Omega$ is its respective angular velocity. The pendulum mass is $m=1\si{\kilogram}$ and the pendulum length is $L=0.5\si{\meter}$. The two variables $\theta$ and $\Omega$ specify the system's state and form a set of state variables of the system.

The system has a stable equilibrium state at $\theta=0,\Omega=0$ and an unstable equilibrium state at $\theta=\pi,\Omega=0$. Suppose that both the initial angle $\theta_0$ and the initial angular velocity $\Omega_0$ are small, one can solve for
\begin{align}
    \theta(t) &= \theta_0 \cos(\omega t) + \frac{\Omega_0}{\omega} \sin(\omega t), \\
    \Omega(t) &= -\omega \theta_0 \sin(\omega t) + \Omega_0 \cos(\omega t),
\end{align}
from the linearized equations of motion, where $\omega=\sqrt{\frac{3g}{2L}}$ is the natural frequency of the system.

To collect the data, we randomly sampled the pendulum arm's initial angle and angular velocity. In total, we collected 1,200 sequences with 60fps. We used 960 of these sequences for training, 120 of them for validation, and 120 of them for testing.

\subsection{Double pendulum}\label{appendix:dataset_double_pendulum}
Our double pendulum system's data was collected from real experiments, using a two colored chaotic pendulum from 3D scientific: the first arm is black and the second arm is blue.
The physical parameters of the double pendulum used for the experiments is shown in  Supplementary~Figure~\ref{fig:dataset_doublependulum}. Using the pivot attachment that came with the pendulum, the pendulum was installed against a brown-beige wall in the laboratory. There are 4 bearings on the pendulum. Three of them are fixed in place and one is left loose to reduce friction. We used an iPhone7 to record videos at 720p and 240fps.

\begin{figure}[H]
\centering
\includegraphics[width=\textwidth]{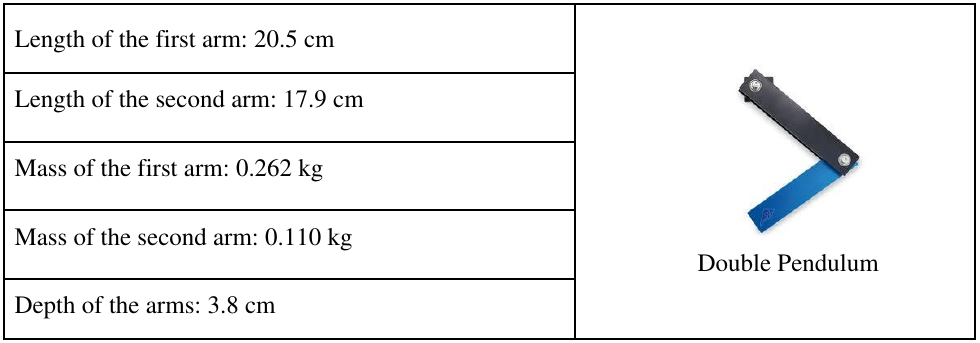}
\caption{Physical parameters of the double pendulum system}\label{fig:dataset_doublependulum}
\end{figure}

While there is no known closed-form, exact solution to the double pendulum system, numerical approximations and simplified analytical solutions for specific cases can be found using methods in classical mechanics. Let us first derive the equations of motion of the double pendulum system, with simplifying assumptions such that the pendulums are 2-dimensional rectangles of uniform density and there are no frictional forces acting on the system.
Denote $m_1$ and $m_2$ the masses of the two arms of the double pendulum, $L_1$ and $W_1$ the length and width of the first arm, and $L_2$ and $W_2$ the length and width of the second arm.
The momenta of inertia of the two arms are:
\begin{align}
I_1 &= \frac{1}{12} m_1 (L_1^2 + W_1^2) \\
I_2 &= \frac{1}{12} m_2 (L_2^2 + W_2^2) . 
\end{align}
We specify the system state by the two arms' angular positions $\theta_1$ and $\theta_2$, and their respective angular velocities $\Omega_1$ and $\Omega_2$.
The kinetic energy of the system is the sum of the two arms' translational and rotational kinetic energies, which is given by:
\begin{align}
T &= \frac{1}{2} \left( \frac{1}{4} m_1 L_1^2 + m_2 L_1^2 + I_1 \right) \Omega_1^2 + \frac{1}{2} \left( \frac{1}{4} m_2 L_2^2 + I_2 \right) \Omega_2^2 + \frac{1}{2} m_2 L_1 L_2 \Omega_1 \Omega_2 \cos (\theta_1 - \theta_2). 
\end{align}
The potential energy of the system is the sum of the two arms' gravitational
potential energies. Taking the configuration that both arms
are horizontal as the zero point, the potential energy of the system is given by:
\begin{align}
V &= - \left( \frac{1}{2} m_1 + m_2 \right) g L_1 \cos \theta_1 - \frac{1}{2} m_2 g L_2 \cos \theta_2. 
\end{align}
Using the Lagrangian $L=T-V$, the equations of motion of the system can be derived as:
\begin{align}
    \dot{\theta}_1 &= \Omega_1 \\
    \dot{\theta}_2 &= \Omega_2, \\
    M_{\theta_1-\theta_2} \begin{bmatrix}
     \dot{\Omega}_1 \\
     \dot{\Omega}_2
     \end{bmatrix} &= \begin{bmatrix}
     - \frac{1}{2} L_1 L_2 \sin (\theta_1 - \theta_2) \Omega_2^2
      - g L_1 \sin \theta_1\\
     \frac{1}{2} L_1 L_2 \sin (\theta_1 - \theta_2) \Omega_1^2
      - \frac{1}{2} g L_2 \sin \theta_2
   \end{bmatrix},
\end{align}
where the matrix
\begin{align}
    M_{\theta_1-\theta_2} &= \begin{bmatrix}
     L_1^2+\frac{I_1}{m_1} & \frac{1}{2} L_1 L_2 \cos (\theta_1 - \theta_2) \\
     \frac{1}{2} L_1 L_2 \cos (\theta_1 - \theta_2) & \frac{1}{4} L_2^2
     + \frac{I_2}{m_2} 
   \end{bmatrix}. 
\end{align}

The total energy of the system is the sum of kinetic and potential energies, which is given by:
\begin{align}
    E &= \frac{1}{2} \left( \frac{1}{4} m_1 L_1^2 + m_2 L_1^2 + I_1 \right) \dot{\theta}_1^2 + \frac{1}{2} \left( \frac{1}{4} m_2 L_2^2 + I_2 \right) \dot{\theta}_2^2 + \frac{1}{2} m_2 L_1 L_2 \dot{\theta}_1 \dot{\theta}_2 \cos (\theta_1 - \theta_2) \notag\\
    &  - \left( \frac{1}{2} m_1 + m_2 \right) g L_1 \cos \theta_1 - \frac{1}{2} m_2 g L_2 \cos \theta_2.
\end{align}

The stable equilibrium state of the system, corresponding to the lowest energy configuration, is at $\theta_1=\theta_2=\Omega_1=\Omega_2=0$. Linearizing the system around this equilibrium state, we obtain the linear system:
\begin{align}
    \dot{\theta}_1 &= \Omega_1, \\
    \dot{\theta}_2 &= \Omega_2, \\
    M_0 \begin{bmatrix}
     \dot{\Omega}_1 \\
     \dot{\Omega}_2
     \end{bmatrix} &= \begin{bmatrix}
      - g L_1 \theta_1\\
      - \frac{1}{2} g L_2 \theta_2
   \end{bmatrix},
\end{align}
where the matrix $M_0$ is given as
\begin{align}
M_0 &= \begin{bmatrix}
     L_1^2+\frac{I_1}{m_1} & \frac{1}{2} L_1 L_2 \\
     \frac{1}{2} L_1 L_2 & \frac{1}{4} L_2^2
     + \frac{I_2}{m_2} 
   \end{bmatrix}. 
\end{align}
By inverting $M_0$, we can write down the linear system as
\begin{align}
    \dot{\theta}_1 &= \Omega_1, \\
    \dot{\theta}_2 &= \Omega_2, \\
    \dot{\Omega}_1 &= -\frac14 gL_1L_2^2 (\theta_1-\theta_2) - \frac{I_2}{m_2} gL_1\theta_1, \\
    \dot{\Omega}_2 &= \frac12 gL_1^2L_2 (\theta_1-\theta_2) - \frac{I_1}{2m_1} gL_2\theta_2.
\end{align}
The solution to this system is a composition of two simple oscillators with frequencies determined by the constants $g,L_1,L_2,m_1,m_2,I_1,I_2$. We numerically evaluated the frequencies as $\omega_1 = 14.007$ and $\omega_2 = 6.404$. 

To construct the dataset, we collected a total of 100 videos,  with an approximate length of 15 seconds for each video. We used 80 of these videos for training, 10 of them for validation, and 10 of them for testing. For better video quality, we trimmed each video to 11s in order to avoid the movement at the beginning and the end of recording caused by humans and small changes in brightness or illumination caused by the camera. Another reason is that  the dynamics towards the late part of the recordings  are more predictable due to the lack of energy and the loss of momentum. Afterwards, we sub-sampled the video to construct a video dataset with 60 fps to produce sufficient visual difference between subsequent frames in a prediction triplet. To feed the video frames into our visual predictive models, the images are resized to 128 $\times$ 128.

Since we are interested in evaluating the results of prediction from the double pendulum system, we further equalized the background of the pendulum system with a simple color filtering so that our vision algorithms can detect the position and orientation of the pendulum arms with another color filtering during the evaluation process. We performed this additional step only to the double pendulum, for the sake of evaluation alone, while other systems do not involve this extra pre-processing step.

\subsection{Cylinder wake}\label{appendix:dataset_cylindrical_flow}
In fluid dynamics, the flow past a circular cylinder is a well-known example that showcases the transition from a laminar steady wake to laminar periodic vortex shedding \cite{duvsek1994numerical,barkley2006linear,noack2003hierarchy}. In the latter regime, vortexes are formed and shed from the cylinder in a regular and oscillating manner, creating a vortex street in the wake. This system is fully described by the two-dimensional incompressible Navier-Stokes equations:
\begin{align}\label{eq:app_ns}
    \frac{\partial\vb{u}}{\partial t} + (\vb{u} \cdot \nabla)\vb{u} &= -\nabla p+\nu\nabla^2\vb{u}, \\
    \nabla \cdot \vb{u} &= 0,
\end{align}
where $\vb{u}=(u_x,u_y)$ represents the flow's velocity field, $p$ represents the pressure divided by the constant density, $\nabla=(\frac{\partial}{\partial x}, \frac{\partial}{\partial y})$ represents the gradient operator, and $\nu>0$ is the kinematic viscosity.

A key quantity for describing the flow patterns is the Reynolds number $Re=U_\infty D_c/\nu$, where $U_\infty$ is the free-stream velocity (velocity far from the cylinder), and $D_c$ is the diameter of the cylinder. It has been demonstrated \cite{duvsek1994numerical,barkley2006linear,noack2003hierarchy} that the solution to \eqref{eq:app_ns} describes a laminar steady wake when the Reynolds number $Re$ is small and it describes laminar periodic vortex shedding when $Re$ is large enough.
 
We generated our dataset by running direct numerical simulations \cite{taira2007immersed,colonius2008fast} on a domain $-15\leq x\leq 35,\,-15\leq y\leq 15$, with a unit-diameter cylinder centered at $x=0,\,y=0$. We set the initial velocity $\vb{u}^0=(u_x^0,0)$ and sample $u_x^0$ from $[0,1]$. Thus, our dataset contains videos for flows with Reynolds numbers $Re$ ranging from 0 to 100. Figure~7 shows the plots for trajectories in smooth neural state variables for different Reynolds numbers (in other words, different initial $u_x$ values).

As demonstrated in \cite{barkley2006linear}, the system has non-unique equilibrium states, and those equilibrium states are all stable when the Reynolds number is small.

\subsection{Computer vision algorithms for estimating physical quantities}

In this section, we provide details on the computer vision algorithms used to estimate physical quantities from images for the four systems under study.

\paragraph{Spring mass position}

\begin{algorithm}[H]
\begin{algorithmic}[1]
\Require $X$ (Spring Mass Image),
$S$ (scale mapping pixels to meters),
$(x_c, y_c)$ (pixel coordinate of center of image)
\Ensure position: spring mass displacement (meters)
\State $(c_x,c_y) \Leftarrow \textbf{rect}(X)$ center of fitted rectangle of mass from image
\State position $\Leftarrow S * (c_x - x_c)$ displacement from the center of the image scaled to meters
\end{algorithmic}\caption{Spring mass position estimation}\label{algo:spring_mass_position}
\end{algorithm}

\paragraph{Single pendulum angle}

\begin{algorithm}[H]
\begin{algorithmic}[1]
\Require $X$ (Single Pendulum Image),
$(x_c, y_c)$ (pixel coordinate of center of image) 
\Ensure angle: single pendulum angle (degrees)
\State $(c_x,c_y) \Leftarrow \textbf{rect}(X)$ center of fitted rectangle of pendulum from image
\State $b\Leftarrow \textbf{box}(X)$ fitted box points of pendulum from image
\State $v = (v_x, v_y) \Leftarrow \textbf{LongEdge}(b)$ vector connecting box corners of longer edge
\State $v_{ref} \Leftarrow (c_x - x_c, c_y - y_c)$ reference vector
\If{ $v \cdot v_{ref} < 0$}
\State $v \Leftarrow -v$ 
\EndIf
\State angle $\Leftarrow \textbf{rad2Angle}(\arctan(\frac{-v_y}{ v_x}) + \frac{\pi}{2})$ counter clockwise angle from downward direction between 0 and 360
\end{algorithmic}\caption{Single pendulum angle estimation}\label{algo:single_pendulum_angle}
\end{algorithm}

\paragraph{Double pendulum angles}

\begin{algorithm}[H]
\begin{algorithmic}[1]
\Require $X$ (Double Pendulum Image), 
$(x_c, y_c)$ (pixel coordinate of center of image) 
\Ensure angle 1: upper pendulum angle (degrees),
angle 2: lower pendulum angle (degrees)
\State $(c^u_x,c^u_y) \Leftarrow \textbf{rect upper}(X)$ center of fitted rectangle of upper pendulum from image
\State $b^u\Leftarrow \textbf{box upper}(X)$ fitted box points of upper pendulum from image
\State $v^u = (v^u_x, v^u_y) \Leftarrow \textbf{LongEdge}(b^u)$ vector connecting box corners of longer edge
\State $v^u_{ref}  \Leftarrow (c^u_x - x_c, c^u_y - y_c)$ reference vector for the upper pendulum
\If{ $v^u \cdot v^u_{ref} < 0$}
\State $v^u \Leftarrow -v^u$ 
\EndIf
\State angle 1$\Leftarrow \textbf{rad2Angle}(\arctan(\frac{-v^u_y}{ v^u_x}) + \frac{\pi}{2})$ counter clockwise angle from downward direction between 0 and 360
\State $(c^l_x,c^l_y)\Leftarrow \textbf{rect lower}(X)$ center of fitted rectangle of lower pendulum from image
\State $b^l\Leftarrow \textbf{box lower}(X)$ fitted box points of lower pendulum from image
\State $v^l = (v^l_x, v^l_y) \Leftarrow \textbf{LongEdge}(b^l)$ vector connecting box corners of longer edge
\State $v^l_{ref} \Leftarrow (c^l_x - 2 * c^u_x + x_c, c^l_y - 2* c^u_y + y_c)$ reference vector for lower pendulum
\If{ $v^l \cdot v^l_{ref} < 0$}
\State $v^l \Leftarrow -v^l$ 
\EndIf
\State angle 2$\Leftarrow \textbf{rad2Angle}(\arctan(\frac{-v^l_y}{ v^l_x}) + \frac{\pi}{2})$ counter clockwise angle from downward direction between 0 and 360
\end{algorithmic}\caption{Double pendulum angles estimation}\label{algo:double_pendulum_angles}
\end{algorithm}

\paragraph{Cylinder wake normalized energy}

\begin{algorithm}[H]
\begin{algorithmic}[1]
\Require $X$ (Cylinder Wake Image), 
$\rho_N$ (density such that maximum velocities leads to $E_N = 1$), 
$c2idx$ (pixel color to $[0,1]$ scale),
$v_x^\text{min}$ (minimum x velocity),
$v_x^\text{max}$ (maximum x velocity),
$v_y^\text{min}$ (minimum y velocity),
$v_y^\text{max}$ (maximum y velocity)
\Ensure $E_N$: energy normalized to [0,1]
\State $V_x \Leftarrow v_x^\text{min} + (v_x^\text{max} - v_x^\text{min}) * c2idx(X)$ pixels mapped to x velocities
\State $V_y \Leftarrow v_y^\text{min} + (v_y^\text{max} - v_y^\text{min}) * c2idx(X)$ pixels mapped to y velocities
\State $E_N \Leftarrow \sum \frac{1}{2} \rho (V_x^2 + V_y^2)$
\end{algorithmic}\caption{Cylinder wake normalized energy estimation}\label{algo:cylinder_wake_energy}
\end{algorithm}

\section{Details on results}\label{appendix:result}

\subsection{Intrinsic dimension estimation}\label{appendix:intrinsic_dim}

In Supplementary~Table~\ref{tab:id}, we provide the estimated intrinsic dimensions for the studied systems, compared to their ground truth values. The intrinsic dimension for each system was estimated by averaging the results of the Levina-Bickel's algorithm on 64 and 8192 dimensional latent vectors, using three different random seeds. The standard errors are also shown for each averaged value. We also provide the rounded values of these estimated intrinsic dimensions to their nearest integers.

\begin{table}[h!]
\begin{tabular*}{\textwidth}{@{\extracolsep\fill}cccccc}
\toprule%
\multicolumn{3}{@{}c@{}}{\textbf{Spring Mass}} & \multicolumn{3}{@{}c@{}}{\textbf{Single Pendulum}} \\
Estimated & Rounded & Ground Truth & Estimated & Rounded & Ground Truth \\
\midrule
$1.96 (\pm 0.01)$ & 2 & 2 &$2.10 (\pm 0.01)$  & 2 &  2\\
\midrule
\multicolumn{3}{@{}c@{}}{\textbf{Double Pendulum}} & \multicolumn{3}{@{}c@{}}{\textbf{Cylinder Wake}} \\
Estimated & Rounded & Ground Truth & Estimated & Rounded & Ground Truth \\
\midrule
 $4.39 (\pm 0.05)$ & 4 & 4 & $2.52 (\pm 0.02)$ & 3 & Unknown\\
\botrule
\end{tabular*}
\caption{The estimated intrinsic dimensions for the studied systems}\label{tab:id}
\end{table}

\subsection{Discovery of smooth neural state variables and neural state vector fields}\label{appendix:smooth_nsv}

We demonstrate the long-term prediction stability and accuracy for our smooth neural state variables, where we compare the computer vision based energy estimation values for the decoded images against the estimated values from their ground truth images. The reject ratio shows the ratio of images that the computer vision estimation fails to derive a value, and it is scaled to remove the cases in which the computer vision estimation fails in the ground truth images themselves. In Supplementary~Figure~\ref{fig:accuracy}(a), we show the long term accuracy of the model rollout  and hybrid rollout procedures as described in \cite{chen_automated_2022}. Note that the smooth neural state variables allow similar level of accuracy and stability to non smooth neural state variables. Furthermore, compared to the dim-$64$ and dim-$8192$ prediction schemes, neural state variables, both smooth and non-smooth, allow stable long term predictions with no rejections. In Supplementary~Figure~\ref{fig:accuracy}(b), we show the long term accuracy and stability of our neural state vector field based predictions. Note that we can achieve a better level of accuracy than the rollout methods, with similar level of stability.  In Supplementary~Figure~\ref{fig:accuracy}(c), we show the neural state variable auto-encoders' pixel reconstruction error, highlighting the similar level of accuracy even when trained with smoothness constraints. Finally in Supplementary~Figure~\ref{fig:accuracy}(d), we show the single step pixel reconstruction error of the neural state vector fields trained on smooth and baseline neural state variables. Neural state vector fields trained on smooth neural state variables are consistently more accurate than those trained on baseline neural state variables in the lower dimensional systems, but we see a slight decrease in accuracy for the cylinder wake system.  

\begin{figure}[H]
    \centering
    \includegraphics[width=1\textwidth]{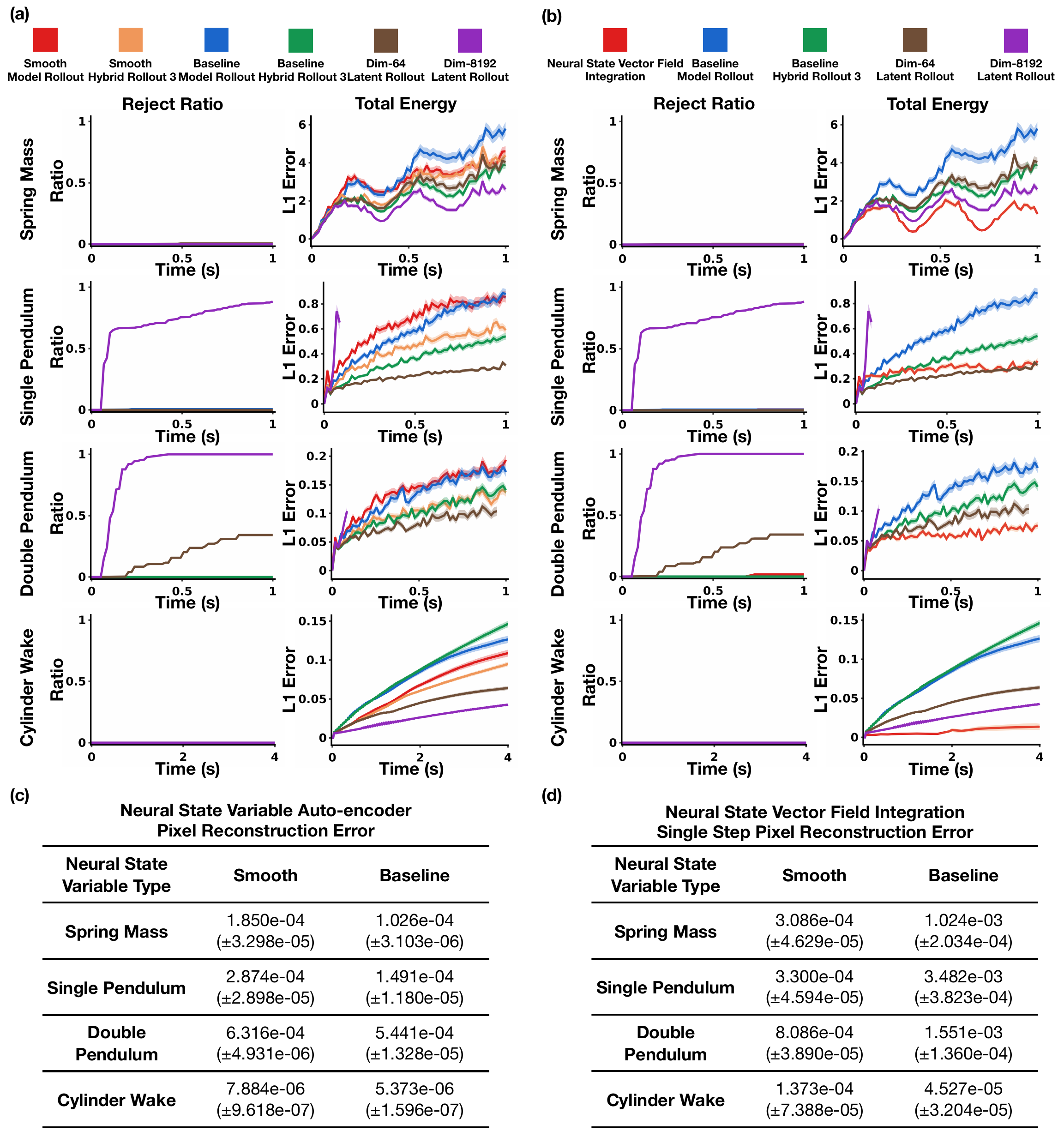}
    \caption{(a) Long-term prediction accuracy from rollout predictions  (b) Long-term prediction accuracy through neural state vector field integration (c) Neural State Variable Auto-encoder Pixel Reconstruction Error (d) Neural State Vector Field Integration Single Step Pixel Reconstruction Error }\label{fig:accuracy}
\end{figure}

While preserving similar levels of predictive accuracy, smooth neural state variables display significantly smoother trajectories. A more detailed comparison of the trajectories shown in Figure 2 are given in  Supplementary~Figure~\ref{fig:trajectory_comparisons}, where we also show the derivative $\frac{\dd\vb{V}_t}{\dd t}$ of each trajectory. We observe that the derivatives are an order of magnitude greater for the baseline neural state variable trajectories compared to our smooth neural state variable trajectories.

\begin{figure}[H]
    \centering
    \includegraphics[width=1\textwidth]{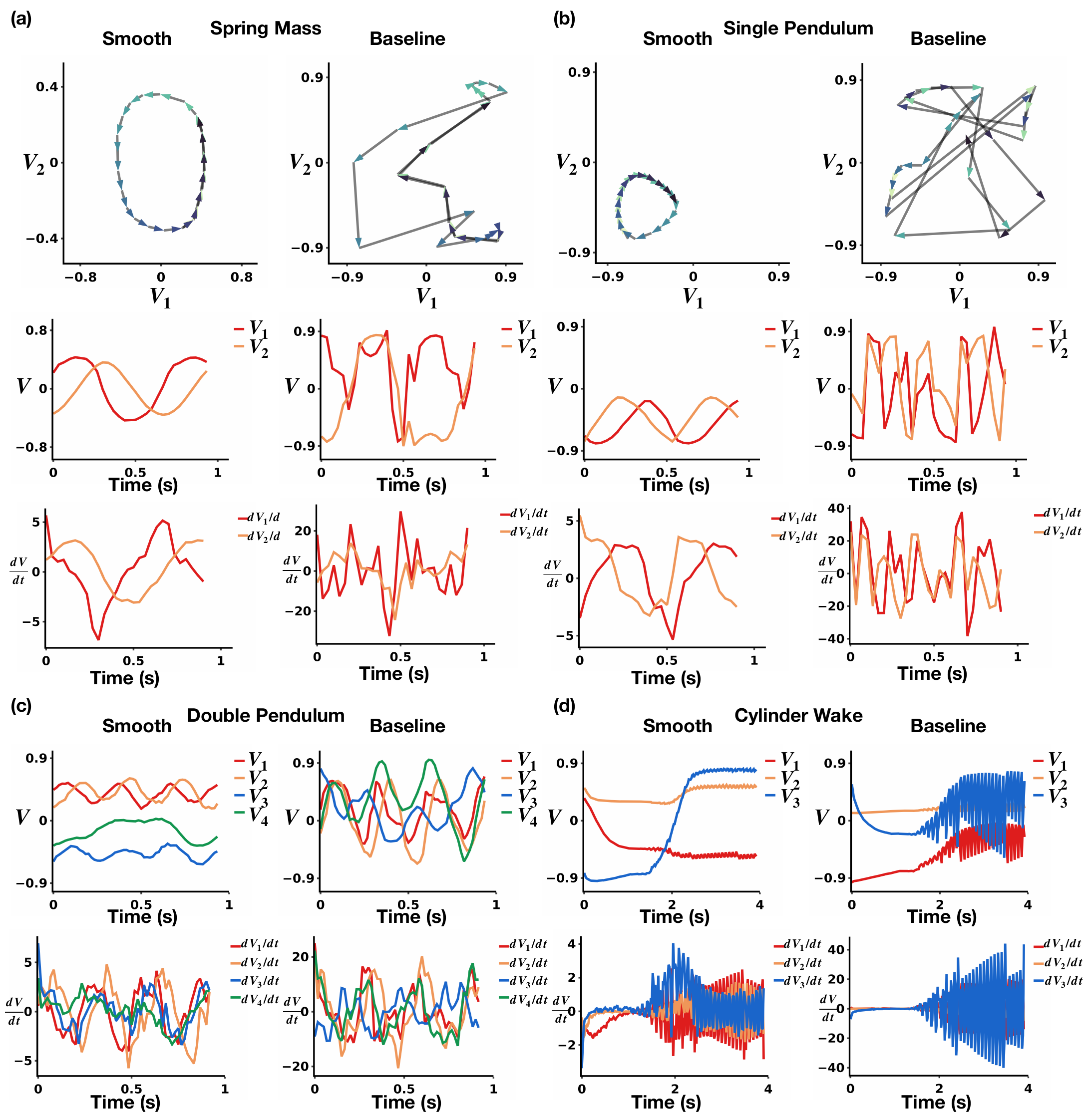}
    \caption{Comparison between smooth and non-smooth neural state variable trajectories. For each system, the same image sequence is encoded to neural state variable trajectories with and without smoothness constraints. Each trajectory is plotted in their respective neural state variable state space as well as a time series against time. We also show the finite difference between neighboring states divided by the time step plotted against time. 
    }\label{fig:trajectory_comparisons}
\end{figure}

To further demonstrate the improved smoothness of our extracted variables, we compared smooth and baseline neural state variables through a family of quantitative smoothness metrics. Let us denote $\vb{V}\in\mathbb{R}^{d}$ the neural state variable with intrinsic dimension $d$. For a continuous-in-time trajectory $\vb{V}(t)$, we can form a generalized metric to measure its smoothness:
\begin{align}
    \mathrm{SM}_{k,p}(\vb{V}(t)) = \begin{cases}
        \left( \int_0^{(N-1)dt} \norm{\frac{d^k\vb{V}(t)}{dt^k}}^p\,dt \right)^{\frac1p}, \quad 1\leq p<\infty, \\
        \max_{t\in[0,(N-1)dt]} \norm{\frac{d^k\vb{V}(t)}{dt^k}}, \quad p=\infty,
    \end{cases} \label{smoothness_metric}
\end{align}
where $k$ is the order of differentiation on $\hat{\vb{V}}$ and $p$ is usually chosen as 1, 2, or $\infty$. As shown in Supplementary~Table~\ref{tab:metric_comparisons}, the results show a consistent trend that our smooth neural state variables exhibit smoother trajectories than the baseline neural state variables. All the above comparisons consistently demonstrate the improved smoothness of our discovered smooth neural state variables compared to the unconstrained baseline neural state variables, and underscore the necessity of enforcing smoothness and the effectiveness of applying smoothness constraints through our method.

\begin{table}[h!]
    \begin{tabular*}{\textwidth}{@{\extracolsep\fill}ccccc}
    \toprule%
    & \multicolumn{2}{c}{\textbf{Spring Mass}} & \multicolumn{2}{c}{\textbf{Single Pendulum}} \\
    \toprule
     & \textbf{Smooth} & \textbf{Baseline} & \textbf{Smooth} & \textbf{Baseline}\\
    \toprule
    $\boldsymbol{\mathrm{SM}_{1,1}}$ & $3.89 (\pm0.05)$ & $5.69 (\pm0.10)$ & $2.10 (\pm0.04)$ & $5.70 (\pm0.13)$ \\
    \midrule
    $\boldsymbol{\mathrm{SM}_{2,1}}$ & $401.36 (\pm22.30)$ & $820.38 (\pm17.83)$ & $136.01 (\pm7.49)$ & $940.81 (\pm30.36)$ \\
    \midrule
    $\boldsymbol{\mathrm{SM}_{1,\infty}}$ & $10.28 (\pm0.31)$ & $36.92 (\pm0.64)$ & $7.41 (\pm0.68)$ & $42.82 (\pm0.95)$ \\
    \midrule
    $\boldsymbol{\mathrm{SM}_{2,\infty}}$ & $1579.85 (\pm92.00)$ & $5214.81 (\pm100.00)$ & $947.09 (\pm96.00)$ & $5977.84 (\pm155.00)$ \\
    \toprule
     & \multicolumn{2}{c}{\textbf{Double Pendulum}} & \multicolumn{2}{c}{\textbf{Cylinder Wake}} \\
    \toprule
     & \textbf{Smooth} & \textbf{Baseline} & \textbf{Baseline} & \textbf{Baseline}\\
    \toprule
    $\boldsymbol{\mathrm{SM}_{1,1}}$ & $3.27 (\pm0.04)$ & $5.86 (\pm0.09)$ & $4.49 (\pm0.12)$ & $31.36 (\pm0.92)$ \\
    \midrule
    $\boldsymbol{\mathrm{SM}_{2,1}}$ & $267.45 (\pm6.82)$ & $417.35 (\pm12.51)$ & $530.40 (\pm17.05)$ & $5021.17 (\pm156.04)$ \\
    \midrule
    $\boldsymbol{\mathrm{SM}_{1,\infty}}$ & $7.41 (\pm0.25)$ & $21.86 (\pm0.51)$ & $2.17 (\pm0.03)$ & $28.23 (\pm0.66)$ \\
    \midrule
    $\boldsymbol{\mathrm{SM}_{2,\infty}}$ & $1134.83 (\pm65.00)$ & $2756.03 (\pm103.00)$ & $368.63 (\pm10.00)$ & $4967.85 (\pm135.00)$ \\
    \bottomrule
    \end{tabular*}
    \caption{Comparison between the calculated average and standard error of the smoothness metrics for smooth and baseline neural state variable trajectories, each averaged over all trajectories in the test data across three random seeds. The derivative norms were further normalized by the range of the neural state variables to ensure a fair comparison.
    }\label{tab:metric_comparisons}
\end{table}

In Supplementary~Figure~\ref{fig:gT_phaseSpace}, the discovered state spaces  are compared to the ground truth phase spaces for the spring mass, single pendulum, and double pendulum systems. This figure demonstrates that our discovered state spaces closely mirror the structure of the ground truth phase spaces and display smooth variation in positions and velocities as states change.

\begin{figure}[H]
    \centering
    \includegraphics[width=1\textwidth]{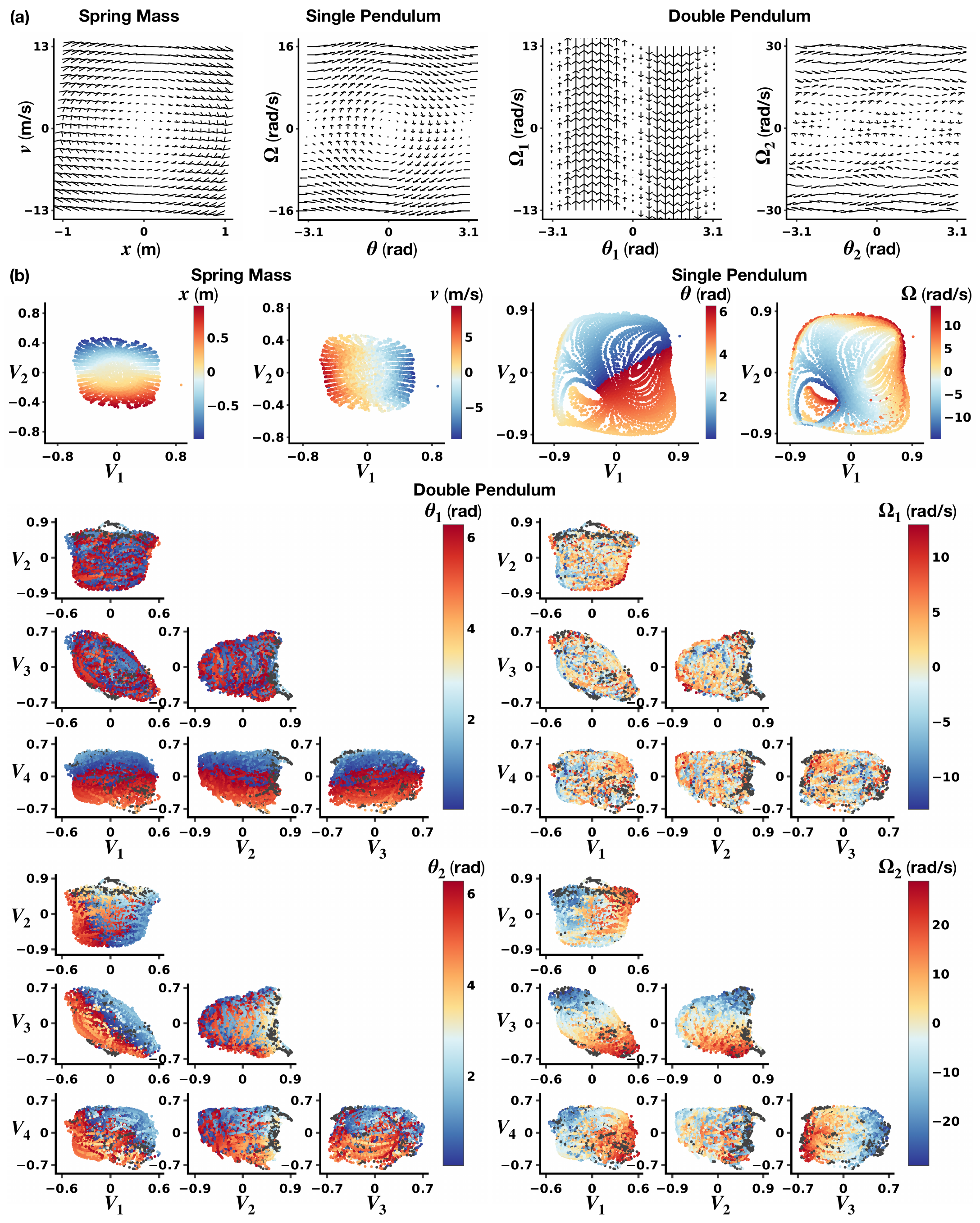}
    \caption{The discovered state space is compared to the ground truth state space. (a) The vector fields for the spring mass, single pendulum, and double pendulum systems (centered around the stable equilibrium) according to classical mechanics are visualized. As the double pendulum system dataset is collected from real observations, the corresponding dynamics are only approximate, and do not account for any non-conservative forces that may exist in the dataset. (b) The discovered neural state variables are visualized against the corresponding physical variables identified by the computer vision algorithms.  }\label{fig:gT_phaseSpace}
\end{figure}

\subsection{Near-equilibrium analysis}\label{appendix:equilibrium_identification}

We describe the equilibrium identification process in further detail in Algorithm~\ref{algo:equilibrium}. Given our set of neural state variables encoded from our test data, the following algorithm returns a set of identified equilibrium states, each labeled with their long term stability. We further show the stability analysis process in Algorithm~\ref{algo:equilibrium_stability}. As per the Lyapunov definition of stability, we used an additional parameter $\mathcal{E}$ for all the $\epsilon$ to test the stability with. For our experiments on the spring mass, single pendulum, and double pendulum systems, we used $C=10$, $n_d=10$, $n_e=10$, $\mathcal{E} = \{ 0.5\%, 1\%, 3\%, 5\%, 10\%\}$, and $T=300$, which is five times longer than the length of the sequences in our datasets. $\Delta t$ used for trajectory integration is the same time interval from our datasets.

\begin{algorithm}[H]
\begin{algorithmic}[1]
\Require $C$ (number of candidate equilibrium states), 
$N$ (number of elements per dimension to perform grid search),
$n_d$ (number of random directions to sample from for each $\epsilon$), $n_e$ (number of sub-samples for each $\epsilon$), $T$ (length of sampled trajectories in number of time steps), $\mathcal{E}$ (list of $\epsilon$ to test for stability)
\Ensure $\mathbf{V}^{\mathrm{eq}}$ the set of identified equilibrium states and their stability
\State $\mathbf{V}^{\mathrm{test}} \Leftarrow \{\mathbf{V}_0, \mathbf{V}_{\Delta t}, \mathbf{V}_{2\Delta t}, \cdots\}$ encoded from sequences in test data
\State $\mathbf{D} \Leftarrow$ the smallest rectangle covering all states in $\mathbf{V}^{\mathrm{test}}$
\State $\mathbf{V}^{\mathrm{candidates}} \Leftarrow$ The $C$ states in $\mathbf{V}^{\mathrm{test}}$ with the lowest $\norm{\hat{F}(\vb{V})}$ values
\If{$\vb{V}$ has dimension $> 2$}
\State $\mathbf{V}^{\mathrm{candidates}} \Leftarrow \ \mathbf{V}^{\mathrm{candidates}} \cup $ $C$ states from $N^d$ uniform grid covering $\mathbf{D}$ with the lowest $\norm{\hat{F}(\vb{V})}$ values
\EndIf
\State $\mathbf{V}^{\mathrm{eq}} \Leftarrow \{\}$
\For{ $\mathbf{V}_{\mathrm{candidate}}$ in $\mathbf{V}^{\mathrm{candidates}}$}
    \If{the root-solving algorithm with $\mathbf{V}_{\mathrm{candidate}}$ as the initial guess successfully returns $\mathbf{V}_{\mathrm{eq}}\in\mathbf{D}$}
    \State stab $\Leftarrow$ Stability($\mathbf{V}_{\mathrm{eq}}$, $n_d$, $n_e$,  $T$, $\mathcal{E}$)
    \State $\mathbf{V}^{\mathrm{eq}} \Leftarrow \mathbf{V}^{\mathrm{eq}} \cup \left\{(\mathbf{V}_{\mathrm{eq}},\ \mathrm{stab})\right\}$
    \EndIf
\EndFor 
\end{algorithmic}\caption{Equilibrium states identification}\label{algo:equilibrium}
\end{algorithm}

The cylinder wake system does not have a unique stable equilibrium state, unlike the other three systems studied. We thus used a different candidate generation process than that shown in Algorithm~\ref{algo:equilibrium} line 3. Instead, we utilized the observed attraction towards an equilibrium state from Section 2.4, and derived a candidate equilibrium state from each trajectory in the test set. These candidate states were generated by taking the average of the final ten states in each trajectory. Figure 4 shows more identified stable equilibrium states for the cylinder wake system. 

\begin{algorithm}[H]
\begin{algorithmic}[1]
\Require $\mathbf{V}_{\mathrm{eq}}$ (identified equilibrium state), $n_d$ (number of random directions to sample from for each $\epsilon$), $n_e$ (number of sub-samples for each $\epsilon$), $T$ (length of sampled trajectories in number of time steps), $\mathcal{E}$ (list of $\epsilon$ to test for stability) 
\Ensure stable: True ($\mathbf{V}_{\mathrm{eq}}$ is a stable equilibrium state) or False ($\mathbf{V}_{\mathrm{eq}}$ is an unstable equilibrium state)
\\
\State $\mathbf{V}^{\mathrm{samples}} \Leftarrow \{\}$
\For{$\epsilon$ in $\mathcal{E}$}
\For{$i=1,2,\cdots,n_d$}
\State $r_i \Leftarrow$ unit vector with random direction
\State $\epsilon_e \Leftarrow \epsilon / n_e$
\For{$j=1,2,\cdots,n_e$}
\State $\vb{V}_{i,j} \Leftarrow \vb{V}_{\mathrm{eq}} + j \times \epsilon_e \times r_i$ 
\State $\vb{V}_{i,j}^{\mathrm{traj}} \Leftarrow$ trajectory by integrating $\hat{F}$ from $\vb{V}_{i,j}$ for $T-1$ steps 
\State $d_{\mathrm{ini}} \Leftarrow \|\vb{V}_i - \vb{V}_{\mathrm{eq}}\| = j \times \epsilon_e$
\State $d_{\mathrm{max}} \Leftarrow \max_{0\leq k< T}\|\vb{V}_{i,j}^{\mathrm{traj}}(k\Delta t) - \vb{V}_{\mathrm{eq}}\|$
\State $\mathbf{V}^{\mathrm{samples}} \Leftarrow \mathbf{V}^{\mathrm{samples}} \cup \left\{(d_{\mathrm{ini}},\,d_{\mathrm{max}})\right\}$
\EndFor
\EndFor 
\EndFor
\State $\mathcal{E}_{stability} \Leftarrow \{ \epsilon : \text{False  } \forall \epsilon \in \mathcal{E}\}$
\For{$\epsilon$ in $\mathcal{E}$}
    \If{ $\exists (d^*_{\mathrm{ini}}, d^*_{\mathrm{max}}) \in \mathbf{V}^{\mathrm{samples}}: d^*_{\mathrm{max}} < \epsilon \land \forall (d_{\mathrm{ini}}, d_{\mathrm{max}}) \in \mathbf{V}^{\mathrm{samples}},\ d_{\mathrm{ini}} \leq d^*_{\mathrm{ini}} \Rightarrow d_{\mathrm{max}} < \epsilon$}
    \State $\mathcal{E}_{stability}[\epsilon] \Leftarrow$ True
    \EndIf
\EndFor
\State stable $\Leftarrow (\forall \epsilon \in \mathcal{E} \ \ \mathcal{E}_{stability}[\epsilon] \leftrightarrow \text{True})$ 
\end{algorithmic}\caption{Stability check}\label{algo:equilibrium_stability}
\end{algorithm}

\subsection{Non-equilibrium analysis}\label{appendix:non_equilibrium}
The rendered videos of the integrated trajectories shown in Figure 5(a) are provided as Supplementary videos. The corresponding filenames for the trajectories of each color are shown in Supplementary~Table~\ref{tab:rendered_videos}. 

\begin{table}[h!]
    \begin{tabular*}{\textwidth}{@{\extracolsep\fill}ccc}
    \toprule%
    & \textbf{Spring Mass} & \\
    \toprule
    \textbf{Purple} & \textbf{Brown} & \textbf{Green} \\
    \toprule
    springMass-traj0.m4v & springMass-traj1.m4v & springMass-traj2.m4v \\
    \toprule
    \textbf{Blue} & \textbf{Yellow} & \textbf{Red}\\
    \toprule
    springMass-traj3.m4v & springMass-traj4.m4v & springMass-traj5.m4v \\
    \toprule%
    & \textbf{Single Pendulum} & \\
    \toprule
    \textbf{Purple} & \textbf{Brown} & \textbf{Green} \\
    \toprule
    singlePendulum-traj0.m4v & singlePendulum-traj1.m4v & singlePendulum-traj2.m4v \\
    \toprule
    \textbf{Blue} & \textbf{Yellow} & \textbf{Red}\\
    \toprule
    singlePendulum-traj3.m4v & singlePendulum-traj4.m4v & singlePendulum-traj5.m4v \\
    \bottomrule
    \end{tabular*}
    \caption{The rendered videos of the integrated trajectories for the spring mass and single pendulum systems shown in Figure 5(a) are included as Supplementary Material with their filenames presented in this table. The colors of the trajectories ordered from smallest to largest amplitudes are: purple, brown, green, blue, yellow, and red.
    }\label{tab:rendered_videos}
\end{table}

The quality of the periodic orbits in the integrated trajectories shown in Figure 5(a) were analyzed in relation to the complexity of the neural network used to approximate the neural state vector field.
The results in Supplementary~Figure~\ref{fig:PO_ablation} show that the complexity of the neural network has little effect on the quality of identified periodic orbits.
While having too few layers may reduce the model's prediction stability, adding additional layers show minimal difference in integrated trajectories from those from our utilized networks. 

\begin{figure}[H]
\begin{center}
\includegraphics[width=1\textwidth]{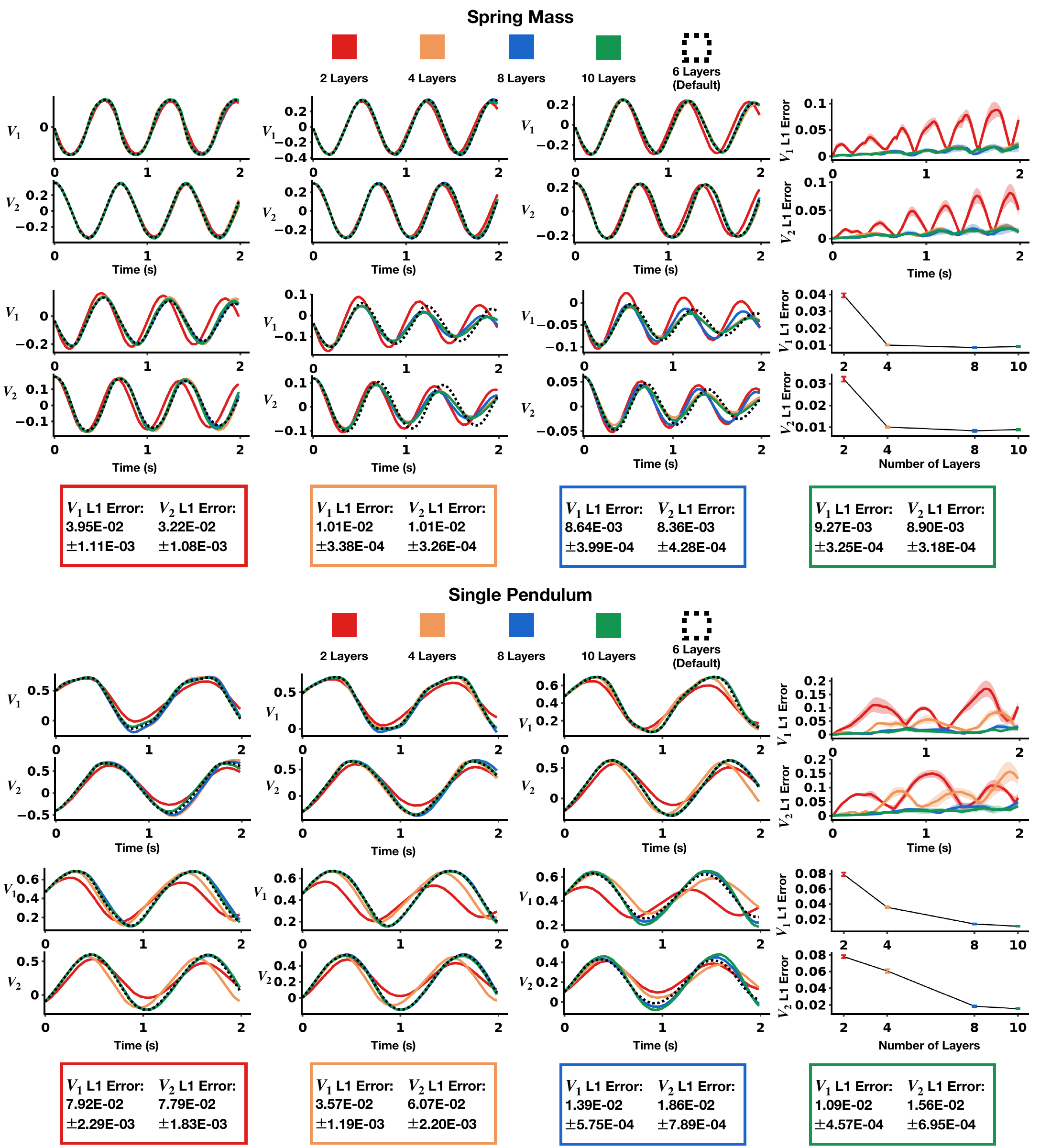}
\caption{Periodic trajectories from identical initial states as in Figure 5(a) are integrated using neural networks of varying depth for the neural state vector field. Deeper neural networks also utilize wider layers with maximum widths increasing in powers of $2$. 
}\label{fig:PO_ablation}
\end{center}
\end{figure}

Although the integrated trajectories do not form perfectly closed orbits, the identified periodic behavior can further be verified by fitting a boundary value problem using our neural state vector field.
In Supplementary~Figure~\ref{fig:periodic_orbit}, the fitted periodic orbits for each trajectory from Figure 5(a) are plotted (solid line) in conjunction to the sampled trajectories (dotted line). The converged periodic orbits are nearly identical to the original integrated trajectories, and we are able to see the similar pattern of uniform period for the spring mass and varying periods for the single pendulum. The mean and margin of error for the L1 errors of the periodic orbits with respect to the sampled trajectories are also plotted below the trajectory plots for each system.

\begin{figure}[H]
    \centering
    \includegraphics[width=1\textwidth]{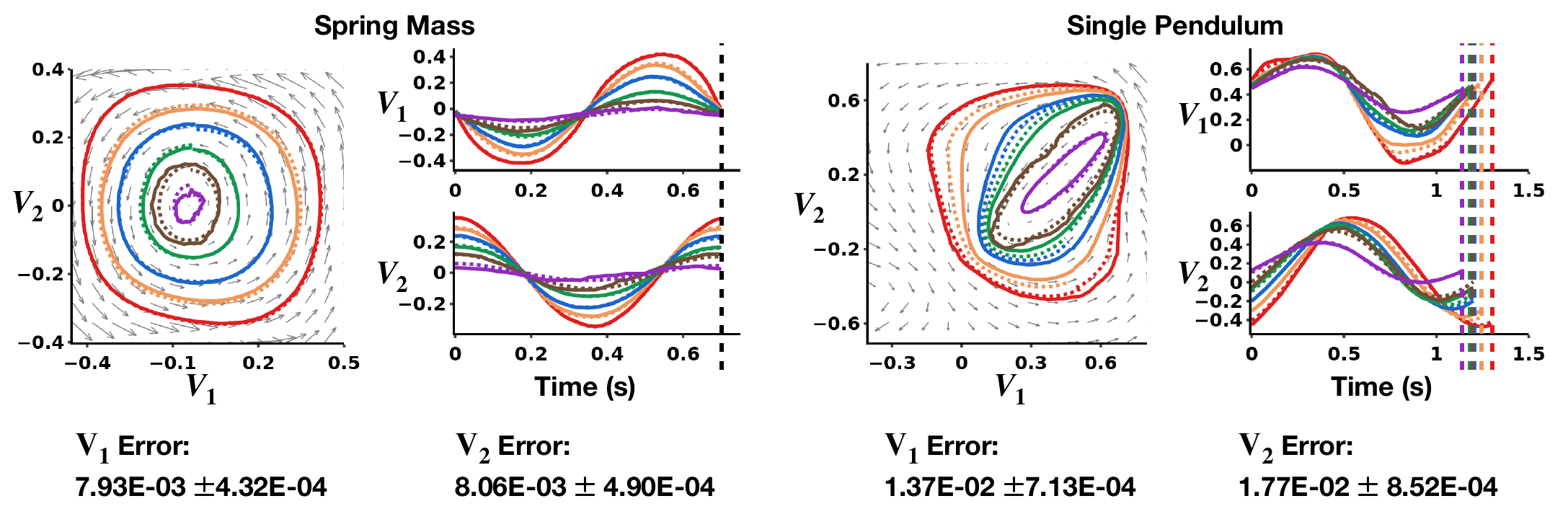}
    \caption{A boundary value problem was solved by fitting an initial guess of one period of the sampled trajectories shown in Figure 5(a). The fitted curves are shown with a solid curve, and the integrated trajectories are shown with a dotted curve of the same color. The length of the period was approximated by finding the first point within the integrated trajectory with the minimum distance to the initial state. The periods are marked with dotted vertical lines. 
    }\label{fig:periodic_orbit}
\end{figure}

\subsection{Synthesizing new data with parameterized  novel dynamics}\label{appendix:new_data_generation}

In this section, we further demonstrate our framework's ability to generate videos that show novel dynamics. Particularly, the novel behaviors are fully controllable through the damping factor $\gamma$ which parameterize the strength of the perturbation in Equation~(4).

\begin{figure}[H]
\centering
\includegraphics[width=\textwidth]{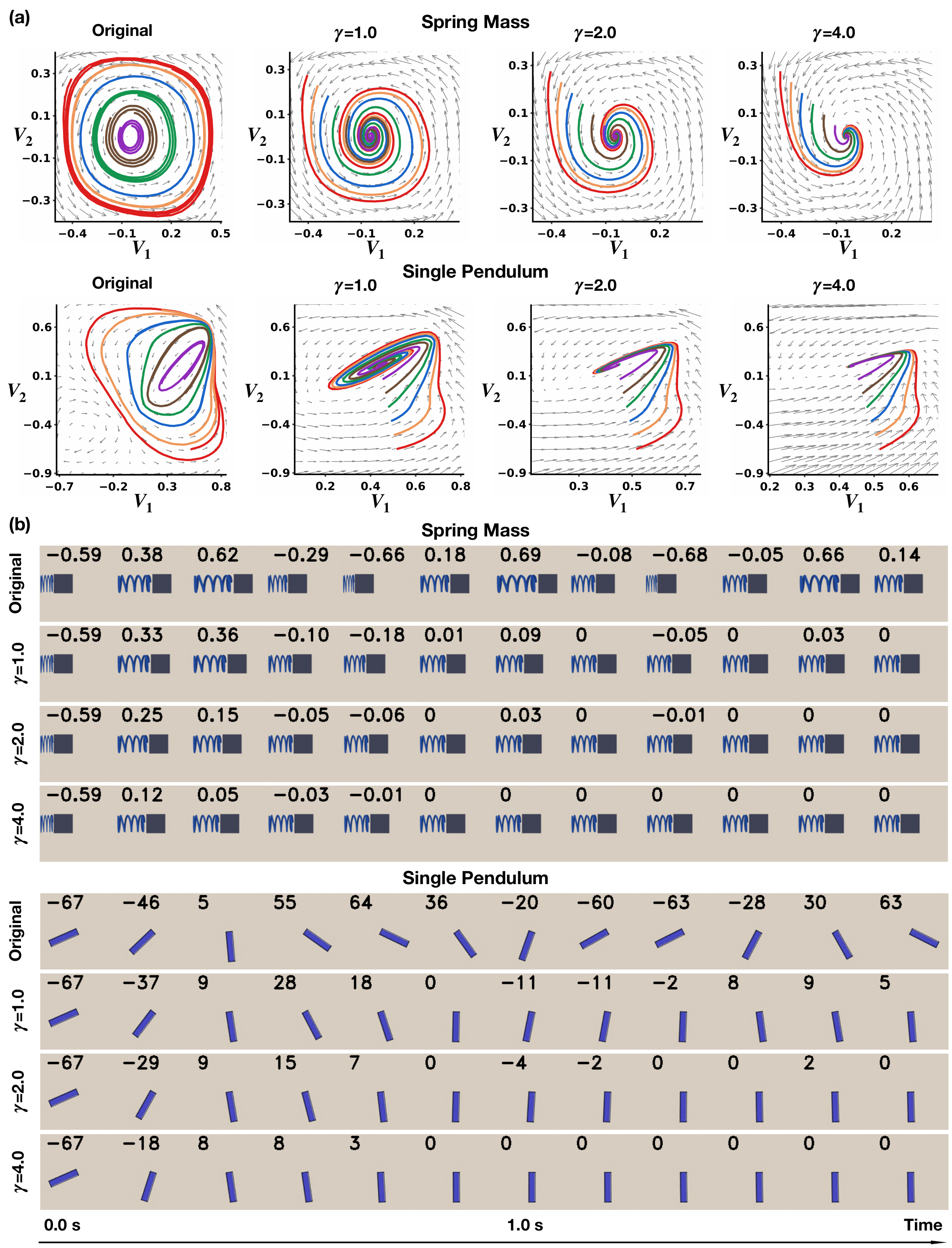}
\caption{Synthesizing new data sequences with parameterized novel dynamics. (a) Plots depict trajectories in the smooth neural state variable space for the spring mass  and single pendulum systems with varying damping factors. (b) Video frames demonstrate the effects corresponding to increasing damping factors. }\label{fig:damping_factor_1}
\end{figure}

Supplementary~Figure~\ref{fig:damping_factor_1}(a) shows trajectories of the new dynamics, as described by Equation (4), in the smooth neural state variable space for the spring mass and single pendulum systems. These trajectories were generated by randomly sampling initial states and integrating the damped system with the damping factor $\gamma$ ranging from zero (original dynamics) to 4.0. The plots reveal that, as the damping factor $\gamma$ increases, the trajectories converge to the equilibrium state more rapidly. The same behavior is exemplified for the double pendulum system in Supplementary~Figure~\ref{fig:damping_factor_2}(a).

In Supplementary~Figure~\ref{fig:damping_factor_1}(b), we present video frames generated from a trajectory of the spring mass and single pendulum systems shown in Supplementary~Figure~\ref{fig:damping_factor_1}(a), sampled at 6 frames per second (fps). The video frames show that while the original dynamics are approximately periodic, the damped dynamics eventually become stationary. Moreover, as the damping factor $\gamma$ is increased, the damping effect becomes more pronounced, causing the system to converge to the stationary state more rapidly. This behavior is similarly observed in the video frames generated from a trajectory of the double pendulum system in Supplementary~Figure~\ref{fig:damping_factor_2}(b).

\begin{figure}[H]
\centering
\includegraphics[width=\textwidth]{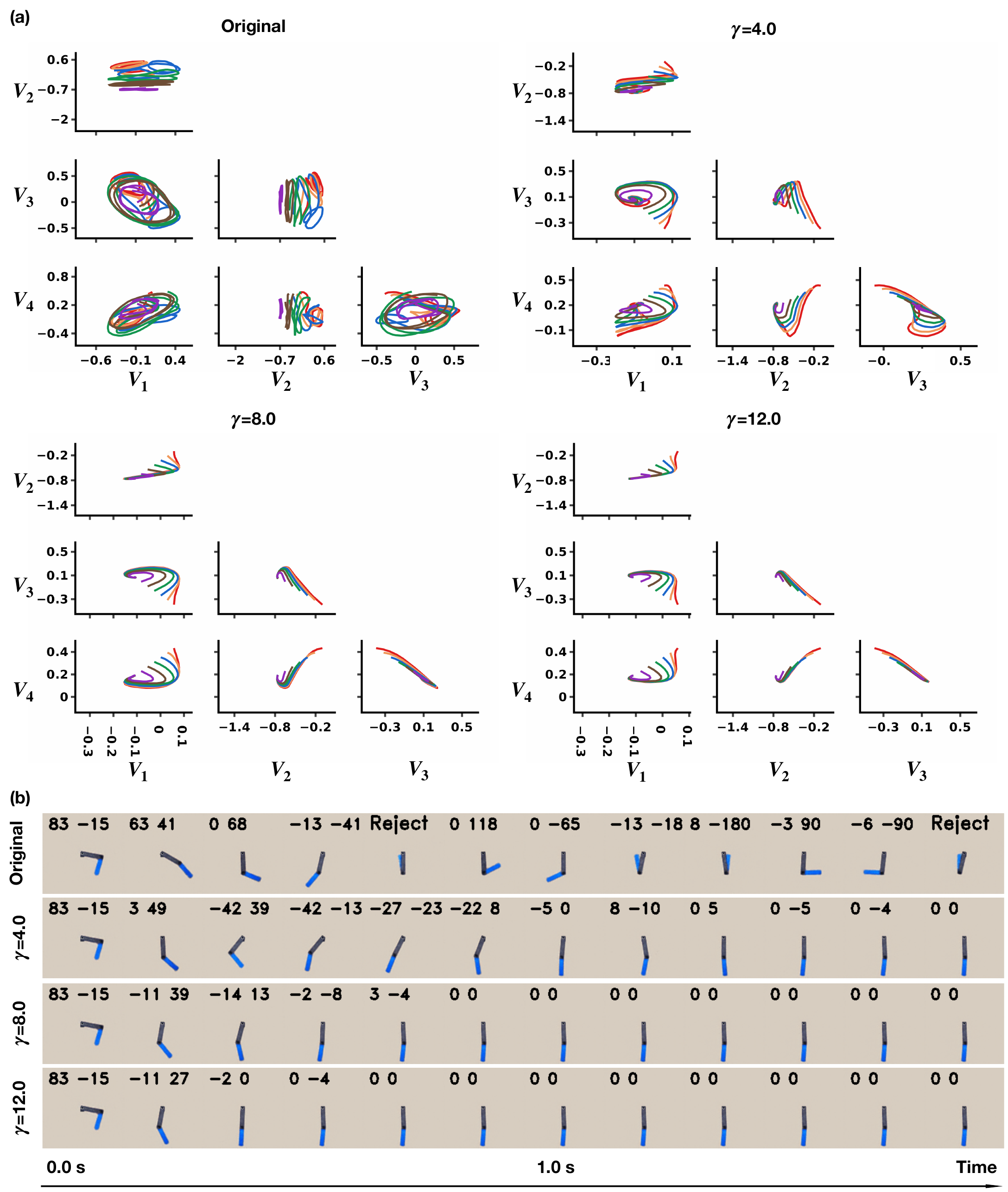}
\caption{Synthesizing new data sequences with parameterized novel dynamics. (a) Plots depict trajectories in the smooth neural state variable space for the double pendulum system with varying damping factors. (b) Video frames demonstrate the effects corresponding to increasing damping factors. }\label{fig:damping_factor_2}
\end{figure}

\subsection{Effects of temporal resolution}

To further analyze the effects of temporal resolution of the provided dataset on model performance, we have run ablation experiments testing our framework on data at a lower temporal resolution (i.\,e., $\Delta t \to 2\Delta t$, $\times0.5\,\text{fps}$). As shown in Supplementary~Figure~\ref{fig:temporalResolution}, our framework achieves comparable predictive accuracy even when data are at a lower temporal resolution, despite velocity information being harder to learn.

\begin{figure}[H]
\begin{center}
\includegraphics[width=1\textwidth]{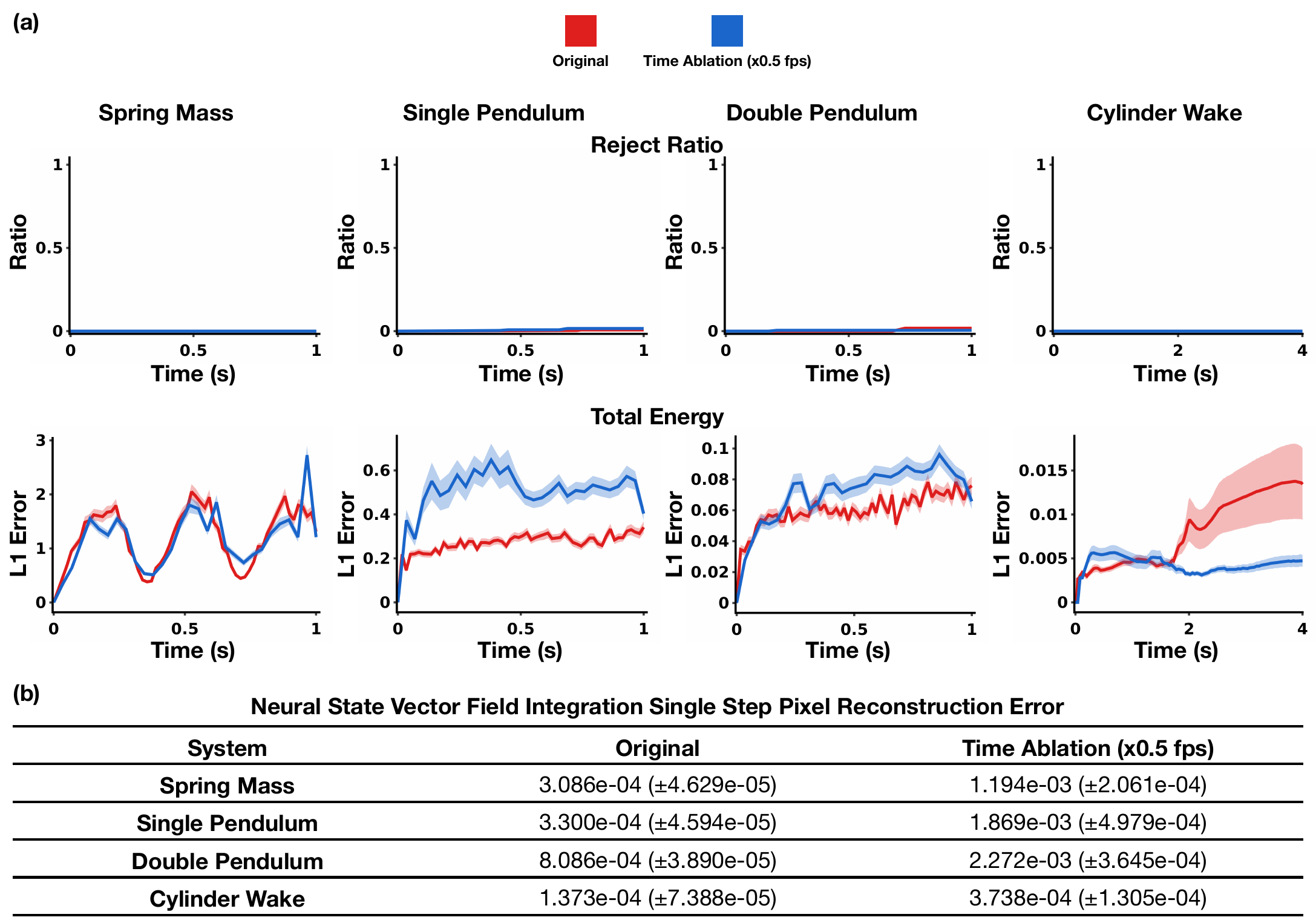}
\caption{Our framework is applied to datasets at a lower time resolution ($\Delta t \to 2\Delta t$, $\times0.5\,\text{fps}$), resulting in comparable accuracies despite the lower temporal resolution. (a) Long-term prediction accuracy through neural state vector field integration; (b) Pixel reconstruction error of a single-step neural state vector field integration.}\label{fig:temporalResolution}
\end{center}
\end{figure}

\section{Implementation details}\label{appendix:implementation_details}

\subsection{Auto-encoders for extracting neural state variables}\label{appendix:model_architecture_g_and_h}

In this section, we provide details about our compression models. We used the same architectures for the auto-encoders $g=(g_E,g_D)$ and $h=(h_E,h_D)$ as implemented in \cite{chen_automated_2022}. 

\paragraph{Auto-encoder $g$}\label{appendix:model_g}

The auto-encoder $g=(g_E,g_D)$ is a neural network with specific parameters listed in Supplementary~Table~\ref{tab:g_architecture_64} (with the latent dimension $N=64$) and Supplementary~Table~\ref{tab:g_architecture_8192} (with the latent dimension $N=8192$).
\begin{table}[h!]
\begin{tabular*}{0.7 \textwidth}{@{\extracolsep\fill}cccccc}
\toprule%
Layer & Kernel Size & $\#$Filters & Stride & Padding & Activation\\
\toprule
Conv1 & $4 \cross 4$ & 32 & 2 & 1 & ReLU\\
\midrule
Conv2 & $4 \cross 4$ & 32 & 2 & 1 & ReLU\\
\midrule
Conv3 & $4 \cross 4$ & 64 & 2 & 1 & ReLU\\
\midrule
Conv4 & $4 \cross 4$ & 64 & 2 & 1 & ReLU\\
\midrule
Conv5 & $4 \cross 4$ & 64 & 2 & 1 & ReLU\\
\midrule
Conv6 & $4 \cross 4$ & 64 & 2 & 1 & ReLU\\
\midrule
Conv7 & $4 \cross 4$ & 64 & 2 & 1 & ReLU\\
\midrule
Conv8 & $3 \cross 4$ & 64 & (1,2) & 1 & ReLU\\
\toprule
Deconv8 & $3 \cross 4$ & 64 & (1,2) & 1 & ReLU\\
\midrule
Deconv7 & $4 \cross 4$ & 64 & 2 & 1 & ReLU\\
\midrule
Deconv6 & $4 \cross 4$ & 64 & 2 & 1 & ReLU\\
\midrule
Deconv5 & $4 \cross 4$ & 64 & 2 & 1 & ReLU\\
\midrule
Deconv4 & $4 \cross 4$ & 64 & 2 & 1 & ReLU\\
\midrule
Deconv3 & $4 \cross 4$ & 32 & 2 & 1 & ReLU\\
\midrule
Deconv2 & $4 \cross 4$ & 16 & 2 & 1 & ReLU\\
\midrule
Deconv1 & $4 \cross 4$ & 3 & 2 & 1 & ReLU\\
\bottomrule
\end{tabular*}
\caption{The architecture of the auto-encoder $g$ with the latent dimension $N=64$}\label{tab:g_architecture_64}
\end{table}
\begin{table}[h!]
\begin{tabular*}{0.7 \textwidth}{@{\extracolsep\fill}cccccc}
\toprule%
Layer & Kernel Size & $\#$Filters & Stride & Padding & Activation\\
\toprule
Conv1 & $4 \cross 4$ & 32 & 2 & 1 & ReLU\\
\midrule
Conv2 & $4 \cross 4$ & 32 & 2 & 1 & ReLU\\
\midrule
Conv3 & $4 \cross 4$ & 64 & 2 & 1 & ReLU\\
\midrule
Conv4 & $4 \cross 4$ & 128 & 2 & 1 & ReLU\\
\midrule
Conv5 & $3 \cross 4$ & 128 & (1,2) & 1 & ReLU\\
\toprule
Deconv5 & $3 \cross 4$ & 64 & (1,2) & 1 & ReLU\\
\midrule
Deconv4 & $4 \cross 4$ & 64 & 2 & 1 & ReLU\\
\midrule
Deconv3 & $4 \cross 4$ & 32 & 2 & 1 & ReLU\\
\midrule
Deconv2 & $4 \cross 4$ & 16 & 2 & 1 & ReLU\\
\midrule
Deconv1 & $4 \cross 4$ & 3 & 2 & 1 & ReLU\\
\bottomrule
\end{tabular*}
\caption{The architecture of the auto-encoder $g$ with the latent dimension $N=8192$}\label{tab:g_architecture_8192}
\end{table}
All convolutional or transposed convolutional layers are accompanied with a batch normalization
layer and a specified activation function. For the encoder network, after each “Conv” layer as
shown in Supplementary~Table~\ref{tab:g_architecture_64} and Supplementary~Table~\ref{tab:g_architecture_8192}, we attached another convolutional layer with the same number of filters as the
current convolutional layer but with 3×3 kernel and 1 as stride. For the decoder network, along
with each “Deconv” layer as shown in Supplementary~Table~\ref{tab:g_architecture_64} and Supplementary~Table~\ref{tab:g_architecture_8192} except for the last one, the input is also passed
through a transposed convolutional layer with the kernel size $4\times4$, 2 as stride, and a Sigmoid
activation function. The output of this branch is then concatenated with each “Deconv”
layer along the feature dimension as the input of the next “Deconv” layer.

All models $g$ in our experiments were trained over 1000 epochs, with an initial learning rate of $0.001$. The learning rate was scaled by $\gamma=0.5$ at the start of epochs 20, 50, 100, and 300. The model from the epoch with the lowest reconstruction error on the validation set was used for subsequent experiments.

\paragraph{Auto-encoder $h$}\label{appendix:model_h}

The auto-encoder $h=(h_E,h_D)$ is also a neural network, with specific parameters listed in Supplementary~Table~\ref{tab:h_architecture}. Each layer is a linear layer accompanied with a sine activation function, and $d$ refers to the system’s intrinsic dimension. The intermediate features whose dimension is $d$ are identified
as our smooth neural state variables. We used the latent vectors produced from the auto-encoder $g$ as outlined in Supplementary~Table~\ref{tab:g_architecture_64} with $N=64$ as the input to $h$.

\begin{table}[h!]
\begin{tabular*}{0.9\textwidth}{@{\extracolsep\fill}cccccc}
\toprule%
Encoder Layer & $\#$Units & Activation & Decoder Layer & $\#$Units & Activation\\
\toprule
Layer1 & 128 & Sine & Layer5 & 32 & Sine\\
\midrule
Layer2 & 64 & Sine & Layer6 & 64 & Sine\\
\midrule
Layer3 & 32 & Sine & Layer7 & 128 & Sine\\
\midrule
Layer4 & $d$ & Sine & Layer8 & 64 & None\\
\bottomrule
\end{tabular*}
\caption{The architecture of the auto-encoder $h$}\label{tab:h_architecture}
\end{table}

All models $h$ in our experiments were trained over 1000 epochs, with an initial learning rate of $0.0005$. The learning rate was scaled by $\gamma=0.5$ at the start of epochs 15, 30, 100, 300, and 500. The model from the epoch with the lowest validation loss was used for subsequent experiments, where the validation loss was calculated using Equation~(5) on the validation set with a fixed $\beta = 1$.

\subsection{Smoothness regularization}\label{appendix:enforcing_smoothness}

In Section 3.2, we introduced Equation~(6), designed to enforce smoothness on the encoded state space by penalizing the neighboring distance between consecutive states. However, without further regularization, the smoothness loss may dominate the the optimization procedure, leading to undesirable effects on the resulting state space. The necessity of the space-filling loss is demonstrated in Supplementary~Figure~\ref{fig:space-filling}, where we show two plots to visualize the distribution of data points over the smooth neural state variable space. The left plot was produced from the auto-encoder $h$ trained with the space-filling loss, while the right one was produced by the auto-encoder $h$ trained without it. When the space-filling loss is included in the loss function, the data points are well spread out, whereas they collapse into a very small region in the smooth neural state variable space if only the smoothness constraint is enforced.

\begin{figure}[H]
    \centering
    \includegraphics[width=\textwidth]{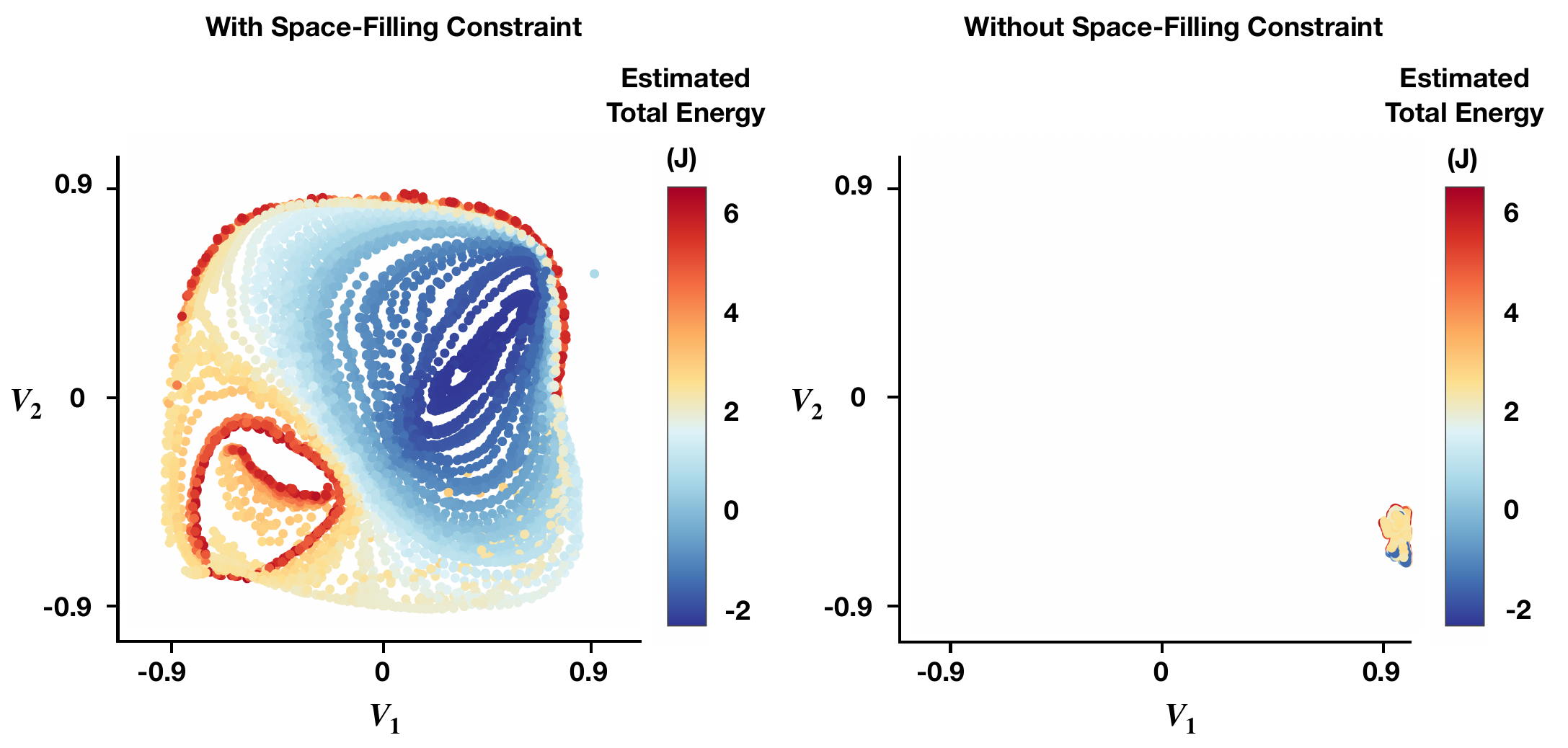}
    \caption{The test data encoded to smooth neural state variables for the single pendulum system, colored by the total energy calculated from computer vision estimated position and velocity from the respective images. When the space-filling loss is not included in the loss function, the embedding space collapses to a small region in the domain, as seen on the right. 
    }\label{fig:space-filling}
\end{figure}

The collapsed state space cannot be utilized to extract accurate dynamics, even when normalization is applied to scale the state space, as demonstrated in Supplementary~Figure~\ref{fig:collapsed_normalized}. A NeuralODE was trained on the normalized state variables, scaled by the standard deviation and biased by the mean of the training data. These results demonstrate that 
both the smoothness loss and space-filling loss are necessary for extracting smooth neural state variables with continuous trajectories that enable learning continuous dynamics via a NeuralODE.

\begin{figure}[H]
\centering
\includegraphics[width=\textwidth]{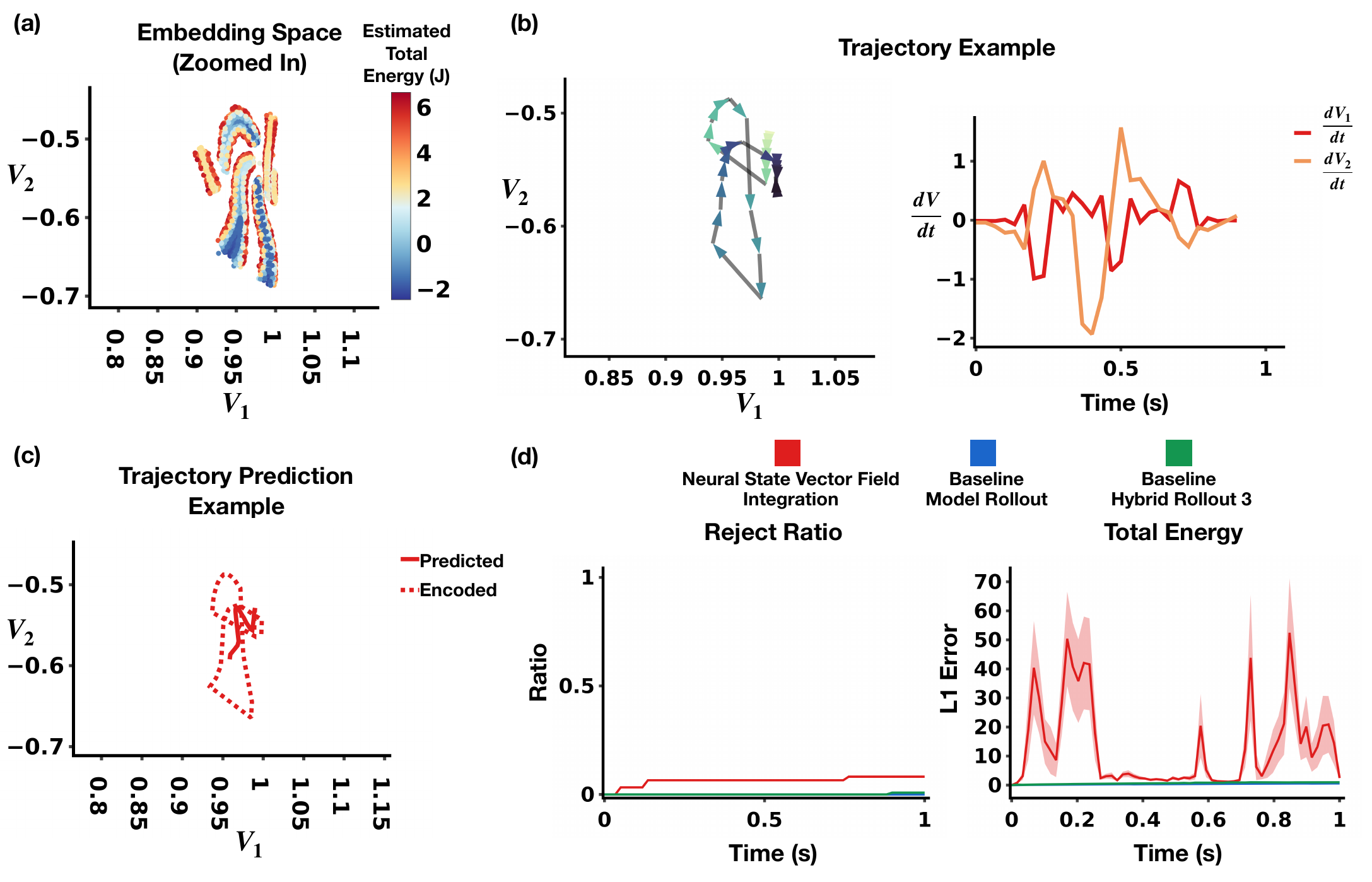}
\caption{Collapsed Space Normalization Results (a) A zoomed in visualization of the collapsed state space. (b) A sample trajectory that is plotted in the zoomed in state space and its finite-difference approximated first order derivatives plotted against time. (c) The predicted trajectory plotted against the example trajectory. (d) Long term prediction error of the trained NeuralODE is worse than even the model rollout prediction scheme from a baseline neural state variable autoencoder.}\label{fig:collapsed_normalized}
\end{figure}

\begin{figure}[H]
    \centering
    \includegraphics[width=1\textwidth]{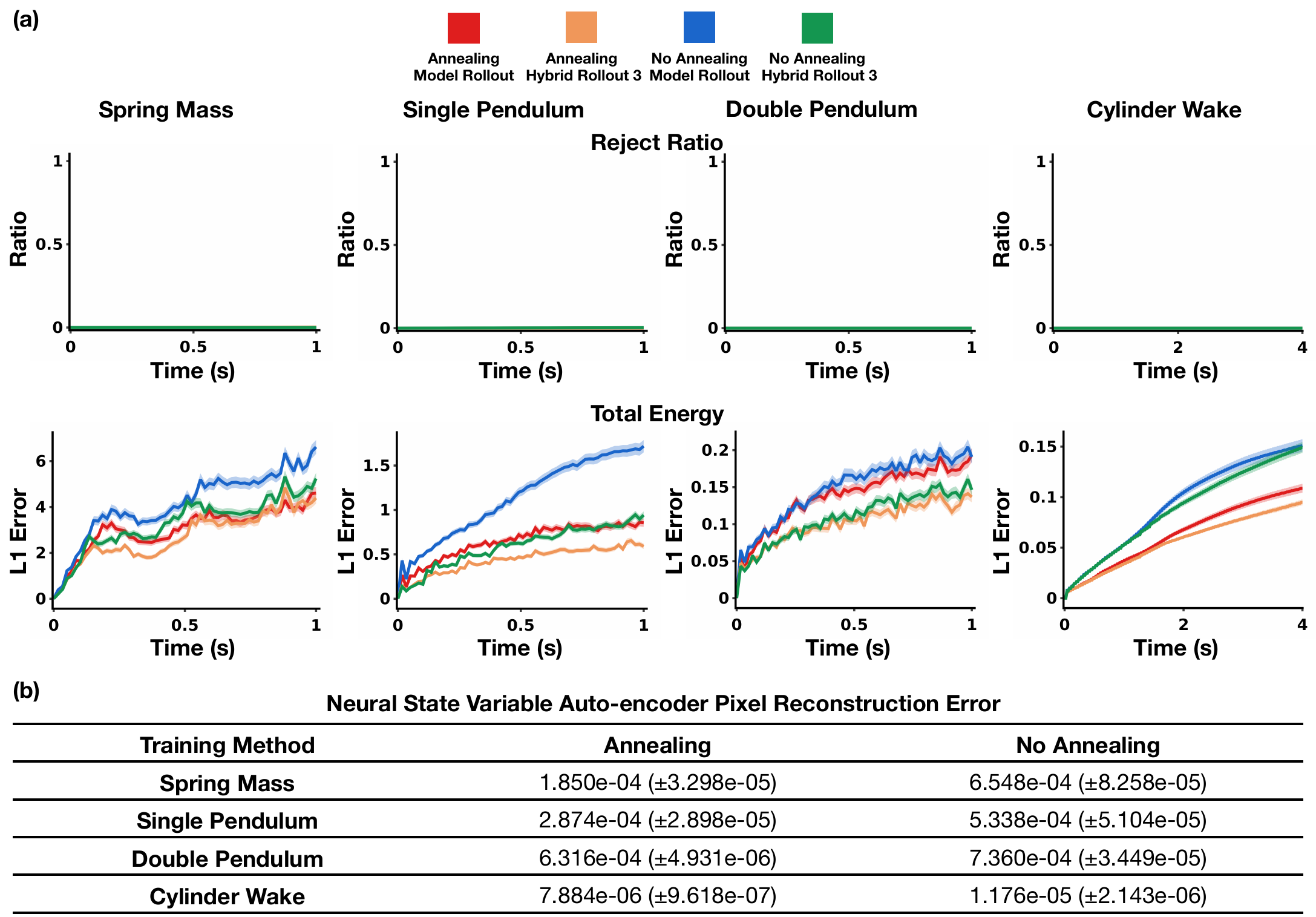}
    \caption{Accuracy Comparison between smooth models trained with and without annealing: (a) Long Term Prediction Stability and Accuracy for Smooth Neural State Variables trained with and without annealing. (b) Auto-encoder Pixel reconstruction error mean and standard error bounds for smooth neural state variable models trained with and without annealing.
    }\label{fig:noAnnealing}
\end{figure}

To reduce the effects of additional constraints on the reconstruction accuracy, we introduced an annealing method during training. We cyclically increased and decreased the weights of the smoothness and space-filling constraints by altering $\beta$ of Equation~(5). We utilized an annealing schedule of 5 cycles with each cycle spanning 200 epochs, where $\beta$ in Equation~(5) was linearly increased from 0 to 1 over the first 100 epochs and kept at 1 for the next 100 epochs. Supplementary~Figure~\ref{fig:noAnnealing} demonstrates the improved accuracy when annealing is applied during training. Supplementary~Figure~\ref{fig:noAnnealing}(a) shows the long-term stability and accuracy for the smooth neural state variable rollout predictions, where we compared the computer vision based energy estimation values for the decoded images against those estimated from the ground truth images. Supplementary~Figure~\ref{fig:noAnnealing}(b) shows the auto-encoder pixel reconstruction mean square error of the smooth neural state variables with their respective standard error bounds. Smooth neural state variables trained with annealing show generally better reconstruction accuracy.

The effects of the smoothness and space-filling constraints are balanced by tuning the weights of the sub-loss terms $w_{\text{smooth}}$, $w_{\text{space}}$, while keeping $L_0$ fixed to a reasonably small value. In our experiments, we used a fixed value of $L_0=0.05$ and tuned the loss weights by optimizing both the reconstruction accuracy and the smoothness metric as defined in Supplementary~Equation~\eqref{smoothness_metric}. The random seeds and corresponding loss weights utilized for each system are shared in Supplementary~Table~\ref{tab:loss_weights}.

\begin{table}[h!]
    \begin{tabular*}{\textwidth}{@{\extracolsep\fill}cccccccc}
    \toprule%
    & & \textbf{Spring Mass} & & & & \textbf{Single Pendulum} &\\
    \toprule
     Random Seed & $1$ & $3$ & $4$ & & $1$ & $2$ & $3$\\
    \midrule
    $w_{\text{smooth}}$ & $64$ & $64$ & $64$ & & $64$ & $128$ & $64$\\
    \midrule
    $w_{\text{space}}$ & $16$ & $8$ & $16$ & & $32$ & $128$ & $64$\\
    \toprule%
    & & \textbf{Double Pendulum} & & & & \textbf{Cylinder Wake} &\\
    \toprule
     Random Seed & $1$ & $2$ & $3$ & & $1$ & $2$ & $3$\\
    \midrule
    $w_{\text{smooth}}$ & $16$ & $16$ & $16$ & & $1024$ & $1024$ & $1024$\\
    \midrule
    $w_{\text{space}}$ & $4$ & $8$ & $8$ & & $0.5$ & $0.5$ & $0.5$\\
    \bottomrule
    \end{tabular*}
    \caption{Loss weights used in our experiments. All values were tuned to optimize the smoothness metric as introduced in Supplementary~Equation~\eqref{smoothness_metric} along with the reconstruction accuracy.
    }\label{tab:loss_weights}
\end{table}

\newpage
\subsection{Implementation of neural state vector fields}\label{appendix:learning_nsvf}

The neural state vector field $\hat{F}$ is implemented as a multi-layer perceptron (MLP), with specific parameters listed in Supplementary~Table~\ref{tab:F_architecture} (with 6 layers) and Supplementary~Table~\ref{tab:F_architecture_deeper} (with 8 layers), where the number of layers were chosen based on the intrinsic dimension $d$ of the system (e.g. 6 layers when $d \leq 2$ and 8 layers when $d > 2$). For both MLPs, each layer is accompanied with the ReLU activation function, except for the last layer.

\begin{table}[h!]
\begin{tabular*}{0.7\textwidth}{@{\extracolsep\fill}ccc}
\toprule%
Layer & $\#$Units & Activation \\
\toprule
Layer1 & 32 & ReLU \\
\midrule
Layer2 & 64 & ReLU \\
\midrule
Layer3 & 128 & ReLU \\
\midrule
Layer4 & 64 & ReLU \\
\midrule
Layer5 & 32 & ReLU \\
\midrule
Layer6 & $d$ & None \\
\bottomrule
\end{tabular*}
\caption{The architecture of the neural state vector field $\hat{F}$}\label{tab:F_architecture}
\end{table}

\begin{table}[h!]
\begin{tabular*}{0.7\textwidth}{@{\extracolsep\fill}ccc}
\toprule%
Layer & $\#$Units & Activation \\
\toprule
Layer1 & 32 & ReLU \\
\midrule
Layer2 & 64 & ReLU \\
\midrule
Layer3 & 128 & ReLU \\
\midrule
Layer4 & 256 & ReLU \\
\midrule
Layer5 & 128 & ReLU \\
\midrule
Layer6 & 64 & ReLU \\
\midrule
Layer7 & 32 & ReLU \\
\midrule
Layer8 & $d$ & None \\
\bottomrule
\end{tabular*}
\caption{The architecture of the neural state vector field $\hat{F}$}\label{tab:F_architecture_deeper}
\end{table}

All models $\hat{F}$ in our experiments were trained over 1000 epochs, with an initial learning rate of $0.0003$. The learning rate was scaled by $\gamma=0.5$ at the start of epochs 15, 30, 100, 300, and 500. The model from the epoch with the lowest validation loss was used for subsequent experiments, where the validation loss was calculated using Equation~(7) on the validation set with a fixed $\rho=0.5$.

\begin{figure}[H]
    \centering
    \includegraphics[width=1\textwidth]{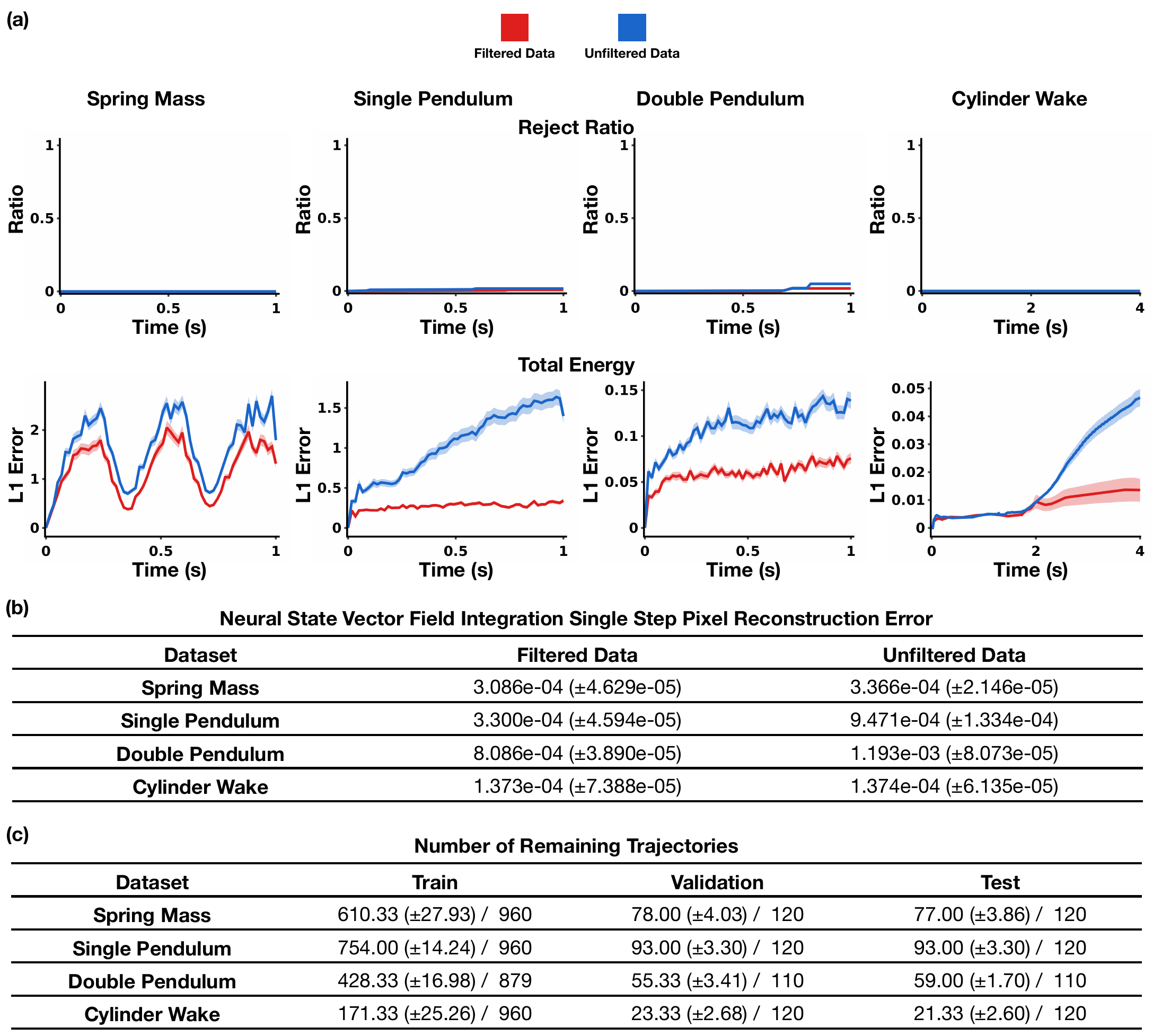}
    ~
    \caption{Accuracy Comparison between neural state vector fields trained with and without filtering: (a) Long Term Prediction Stability and Accuracy for Neural State Vector Fields trained with and without filtering. (b) Neural State Vector Field integration single step pixel reconstruction error mean and standard error bounds when trained and tested on filtered and unfiltered data. (c) Average number of trajectories remaining after filtering, reported with standard error bounds.
    }\label{fig:noFilter}
\end{figure}

To improve the accuracy of the trained neural state vector field, we applied filtering to our neural state variable trajectories before training. We removed data sequences that contain trajectories where the distance between any consecutive state is greater than the $99$~-percentile distribution within the respective dataset subset 
(e.g. training, validation or test).  Supplementary~Figure~\ref{fig:noFilter} demonstrates the improved accuracy when filtering is applied before training the neural state vector field.  Supplementary~Figure~\ref{fig:noFilter}(a) shows the long-term prediction stability and accuracy for the neural state vector field integrated neural state variable trajectories, where we compared the computer vision based energy estimation values for the decoded images against the energy estimated from the ground truth images.  Supplementary~Figure~\ref{fig:noFilter}(b) shows the single step pixel reconstruction mean square error of the neural state vector field predictions with their respective standard error bounds. Using filtered trajectories shows consistent improvement in both cases.  Supplementary~Figure~\ref{fig:noFilter}(c) shows the remaining number of trajectories after filtering for the training, validation and test sets, averaged over three random seeds along with their respective standard error bounds.

\begin{figure}[H]
    \centering
    \includegraphics[width=1\textwidth]{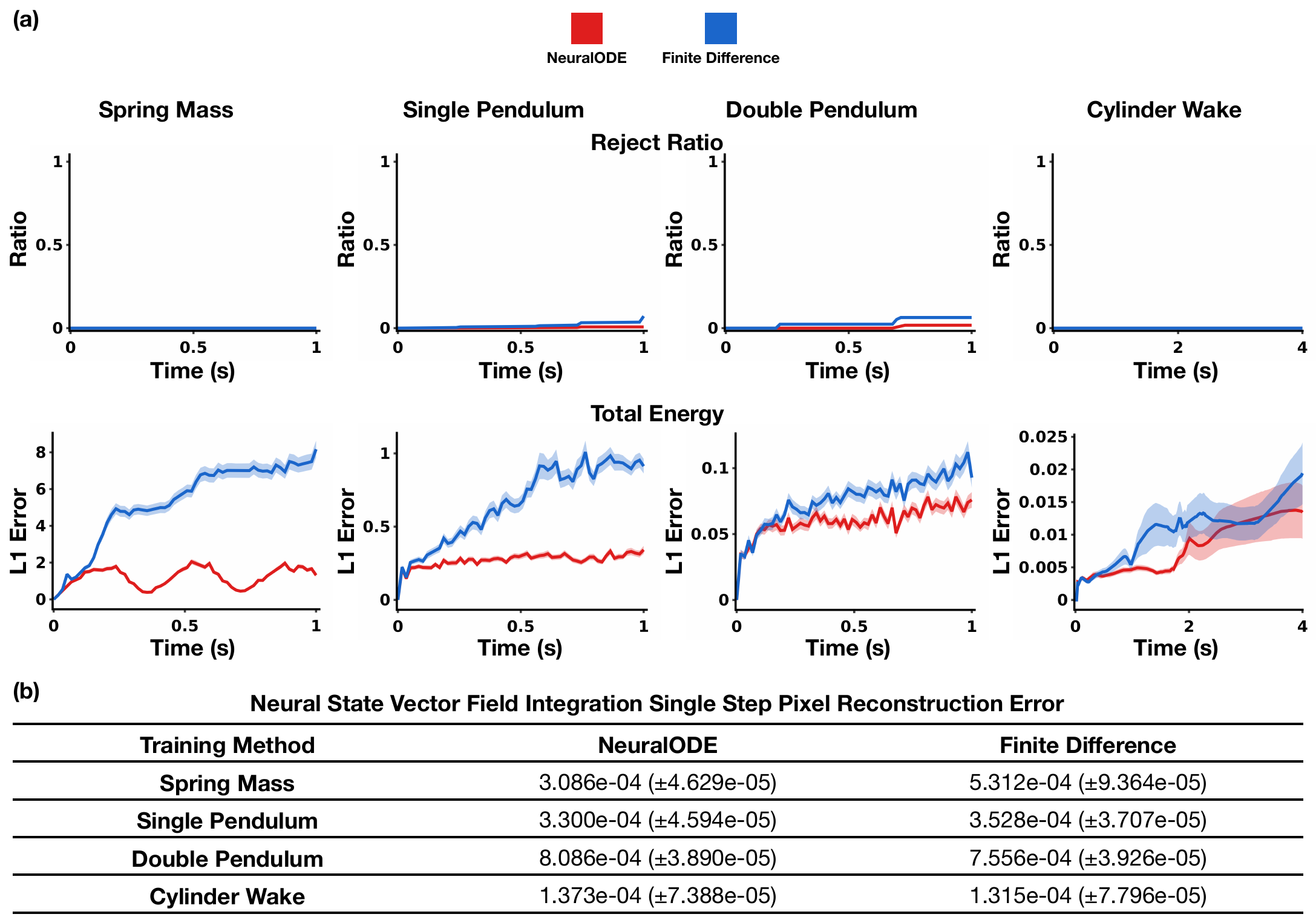}
    ~
    \caption{Accuracy Comparison between neural state vector fields trained as a NeuralODE and neural state vector fields trained to reconstruct the finite difference between consecutive states: (a) Long Term Prediction Stability and Accuracy for Neural State Vector Fields trained as neuralODE and Neural State Vector Fields trained to reconstruct the approximated rate of change using finite difference. (b) Neural State Vector Field integration single step pixel reconstruction error mean and standard error  bounds when trained as a NeuralODE compared to when trained to reconstruct the finite difference between consecutive states.
    }\label{fig:discrete}
\end{figure}

Furthermore, we implemented the training procedure such that the neural state vector field is trained as a NeuralODE \cite{chen2019neuralordinarydifferentialequations, politorchdyn}, allowing us to optimize for the weighted reconstruction accuracy over longer time horizons, rather than explicitly fitting the outputs of the neural state vector field to reproduce the finite difference approximated rate of change for a single step. 
The relative weights of latter states in the trajectories were annealed over 4 cycles of 250 epochs each, where $\rho$ in Equation~(7) was linearly increased from $0.1$ to $0.9$ over the first 125 epochs and kept at $0.9$ over the remaining 125 epochs. 
Supplementary~Figure~\ref{fig:discrete} demonstrates the improved accuracy when utilizing the NeuralODE paradigm for training.  Supplementary~Figure~\ref{fig:discrete}(a) shows the long-term prediction stability and accuracy for the neural state vector field integrated neural state variable trajectories, where we compare the computer vision based energy estimation values for the decoded images against the energy estimated from the ground truth images.  Supplementary~Figure~\ref{fig:discrete}(b) shows the single step pixel reconstruction mean square error of the neural state vector field predictions with their respective standard error bounds. Using the NeuralODE training method shows consistent improvement in long term prediction accuracy.

\end{appendices}

\bibliography{sn-bibliography}

\end{document}